\documentclass[aps,prx,longbibliography,twocolumn,nopacs,superscriptaddress]{revtex4-1}

\usepackage[english]{babel}
\usepackage{graphicx}
\usepackage{amsmath}
\usepackage{amsfonts}
\usepackage{times}
\usepackage{amssymb}
\usepackage{units}
\usepackage{color}
\usepackage{eurosym}
\usepackage[ansinew]{inputenc}
\usepackage[colorinlistoftodos]{todonotes}
\usepackage{hyperref}

\newcommand{\sqthree}{\ensuremath{\sqrt{3}\times\sqrt{3}\,\text{R}30^\circ}}
\newcommand{\BiPbAg}{\ensuremath{\text{Bi}_x\text{Pb}_{1-x}\text{/Ag(111)}}}

\newcommand{\BixPbyAg}[2]{\ensuremath{\text{Bi}_{#1}\text{Pb}_{#2}\text{/Ag(111)}}}

\begin{document}

\title{Visualizing the multifractal wavefunctions of a disordered two-dimensional electron gas}

\author{Berthold J{\" a}ck}
 \affiliation{Max-Planck-Institut f\"ur
  Festk\"orperforschung, 70569 Stuttgart, Germany}
  \affiliation{Princeton University, Joseph Henry Laboratories and Department of Physics, Princeton, NJ 08544, USA}
\author{Fabian Zinser}
 \affiliation{Max-Planck-Institut f\"ur
  Festk\"orperforschung, 70569 Stuttgart, Germany}
\author{Elio J. K{\" o}nig}
 \affiliation{Max-Planck-Institut f\"ur
  Festk\"orperforschung, 70569 Stuttgart, Germany}
 \affiliation{Rutgers University, Department of Physics, Piscataway, New Jersey 08854}  
\author{Sune N.\ P.\ Wissing}
 \affiliation{Physikalisches Institut, Westf\"alische Wilhelms-Universit\"at M\"unster, 48149 M\"unster, Germany}
\author{Anke B.\ Schmidt}
 \affiliation{Physikalisches Institut, Westf\"alische Wilhelms-Universit\"at M\"unster, 48149 M\"unster, Germany}
\author{Markus Donath}
 \affiliation{Physikalisches Institut, Westf\"alische Wilhelms-Universit\"at M\"unster, 48149 M\"unster, Germany}
\author{Klaus Kern}
 \affiliation{Max-Planck-Institut f\"ur
  Festk\"orperforschung, 70569 Stuttgart, Germany}
  \affiliation{Institut de Physique de la Mati{\`e}re Condens{\'e}e, Ecole Polytechnique F{\'e}d{\'e}rale de Lausanne, 1015 Lausanne, Switzerland}
\author {Christian R. Ast}
 \affiliation{Max-Planck-Institut f\"ur
  Festk\"orperforschung, 70569 Stuttgart, Germany}

\date{\today}

\begin{abstract}
The wavefunctions of a disordered two-dimensional electron gas at the quantum-critical Anderson transition are predicted to exhibit multifractal scaling in their real space amplitude. We experimentally investigate the appearance of these characteristics in the  spatially resolved local density of states of the 2D mixed surface alloy Bi$_{\rm x}$Pb$_{\rm(1-x)}$/Ag(111), by combining high-resolution scanning tunneling microscopy with spin- and angle-resolved inverse-photoemission experiments. Our detailed knowledge of the surface alloy's electronic band structure, the exact lattice structure and the atomically resolved local density of states enables us to construct a realistic Anderson tight binding model of the mixed surface alloy, and to directly compare the measured local density of states characteristics with those from  our model calculations. The statistical analyses of these two-dimensional local density of states maps reveal their log-normal distributions and multifractal scaling characteristics of the underlying wavefunctions with a finite anomalous scaling exponent. Finally, our experimental results confirm theoretical predictions of an exact scaling symmetry for Anderson quantum phase transitions in the Wigner-Dyson classes.

\end{abstract}

\maketitle

\section{Introduction}

Multifractality (MF) is an ubiquitous phenomenon in nature that characterizes a variety of physical phenomena from turbulence \cite{sreenivasan1991} to fluid flow and surface growth \cite{stanley1988multifractal}. It describes the power-law scaling of fractal quantities with not one but a continuous (infinite) set of anomalous scaling exponents. MF also governs the physics of the Anderson quantum phase transition in disordered 
systems~\cite{abrahams1979scaling, wegner1979mobility},
which separates phases with Anderson localized~\cite{anderson1958} and extended wave functions in insulators and metals, respectively. In this sense, Anderson transitions are substantially richer than ordinary quantum phase transitions, where the set of anomalous scaling exponents is finite.

\begin{figure}
  \includegraphics[width = 1\columnwidth]{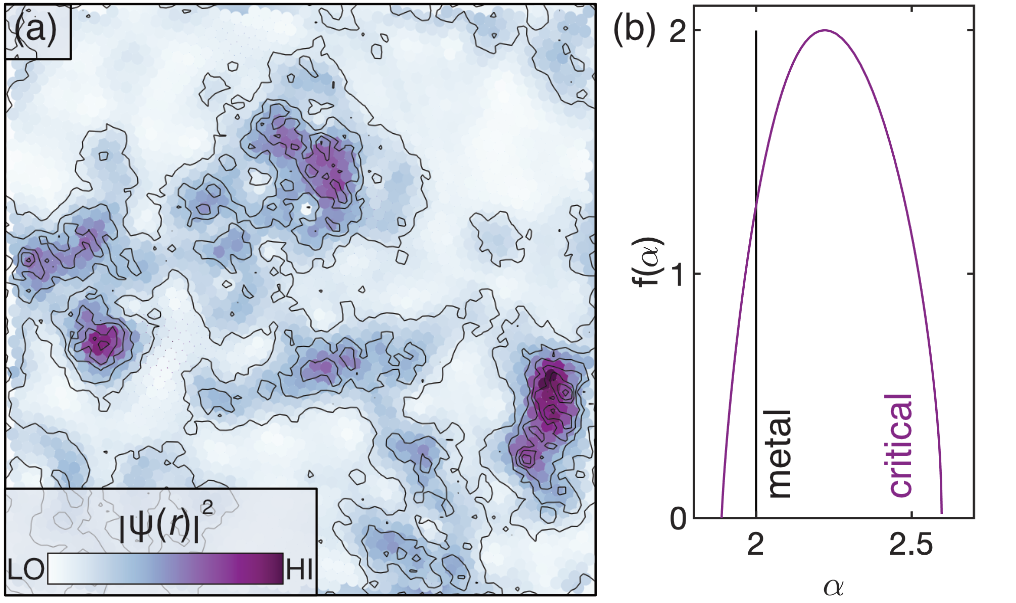}
  \caption{{\bf Multifractal wavefunction scaling in two dimensions.} (a) Calculated LDOS ($|\Psi(\vec{r})|^2$) map (approx. $50\times50$ lattice sites) of a two-dimensional electron gas exhibiting multifractal scaling behavior with power-law decay characteristics. The LDOS scales with different exponents $\alpha$ along different topographic lines (black solid lines). 
  (b) Schematic plot of the singularity spectrum $f(\alpha)$ for a $d=2$ dimensional system with multifractal (purple)and metallic (black) scaling behavior.}
  \label{fig:1}
\end{figure}

At Anderson criticality, MF characterizes the real space amplitude fluctuations of the underlying
quantum mechanical wavefunctions $\Psi(\vec{r})$
\cite{castellani1986multifractal} and thereby the real space distribution of the corresponding local electronic density of states 
(LDOS) \cite{lerner1988distribution, mirlin1994distribution}.
Geometrically, it accounts for the spectrum of fractal dimensions associated with topographic lines in the real space distribution of $|\Psi(\vec{r})|^2$ , as illustrated in Fig.\,\ref{fig:1}(a). Defining such lines by $|\Psi(\vec{r})|^2 = \text{const.}\times L^{-\alpha}$ ($L$ is the system size) \cite{halsey1986fractal, evers2008anderson}
the probability distribution for the observation of a real space LDOS scaling at a given singularity strength $\alpha$ is called the ``singularity spectrum'', $f(\alpha)$. \cite{mandelbrot1974intermittent,hentschel1983infinite, halsey1986fractal}; i.\ e.\ by changing the system size L, $|\Psi(\vec{r})|^2$ scales with the same exponent along one topographic line. In the presence of MF at the Anderson transition, $f(\alpha)$ describes a continuous set of scaling exponents, which, away from the quantum critical point in the metallic phase, collapses onto a line for scaling with a single exponent (Fig.\,\ref{fig:1}(b)).

Scanning Tunneling Microscopy (STM) experiments can probe the LDOS with atomic resolution, and they are particularly useful to study wavefunction multifractality of two-dimensional (2D) systems. Previous STM experiments on disordered two-dimensional quantum Hall systems \cite{morgenstern2003real, hashimoto2008quantum}, on the surface of bulk GaAsMn \cite{richardella2010visualizing}, and on 2D layers of transition metal dichalcogenides \cite{zhao2019disorder} reported $f(\alpha)$-spectra, calculated from differential tunneling conductance, $dI/dV$, maps, whose broadened characteristics were consistent with an interpretation of multifractal scaling at criticality (cf. Fig.\,\ref{fig:1}(b)). 

However, such $dI/dV$-maps do not necessarily reflect the pure LDOS, but commonly contains contribution from other phenomena such as charge density waves \cite{ugeda2016characterization}, quasiparticle interference \cite{zhao2019disorder} and electronic correlation \cite{richardella2010visualizing}, thereby obscuring the underlying eigenstate distribution characteristics. Without a sufficiently large measurement window \cite{morgenstern2003real} or full access to the wavefunction decay in the 3D bulk \cite{richardella2010visualizing}, a complete picture of the eigenstate real space decay cannot be captured either. Since little is known about the underlying local electronic structure in the presence of disorder in most material systems studied to date, experimental results cannot be compared with theoretical expectations, which all together render the interpretation of a multifractal analysis challenging.

In this article, we experimentally investigate the LDOS of the 2D mixed surface alloy Bi$_{x}$Pb$_{1-x}$Ag and report multifractal wavefunction scaling in a disordered 2D electron gas with strong spin-orbit coupling. To this end, we combine high-resolution STM measurements with inverse-photoemission experiments (IPE). This combined approach provides us with detailed knowledge on the local electronic properties of the surface alloy and the atomic scale lattice structure of the disordered surface alloy with chemical contrast. On this basis, we construct a realistic Anderson tight binding model of the disordered surface alloy, which allows to compare the measured $dI/dV$-map characteristics with the calculated LDOS maps. The quantitative analyses of these maps reveal their log-normal distributions and the multifractal scaling of the mixed surface alloy eigenstate density. Finally, our experimental results are in agreement with theoretical expectations of an exact symmetry of the scaling exponents for Anderson quantum phase transitions in the Wigner-Dyson classes.

\section{Results}
\begin{figure*}
  \includegraphics[width = 0.7\textwidth]{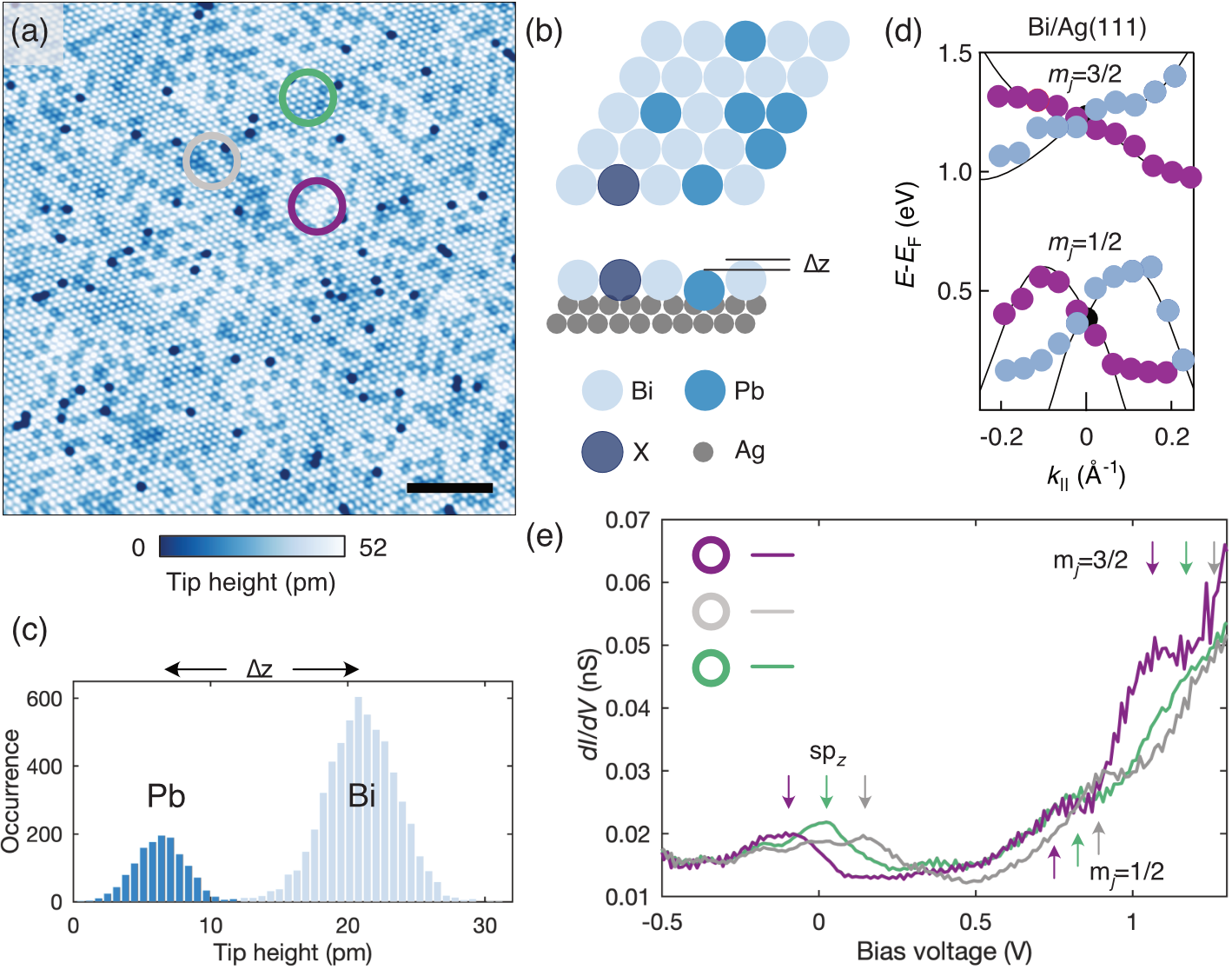}
  \caption{{\bf Electronic properties and atomic lattice structure of the mixed surface alloy.} (a) STM topography of the \BixPbyAg{0.79}{0.21}\ mixed surface alloy (setpoint $V=3\,$mV, $I=7\,$nA). Scale bar 5\,nm. (b) Schematic sketch of the mixed surface alloy's atomic lattice structure. The alloy atoms substitute every third Ag atom of the surface layer, forming a \sqthree lattice structure. $\Delta$z denotes the apparent height difference between the Bi and Pb atoms as measured with the STM; 'X' denotes an alloy lattice vacancy. (c) Histogram analysis of the topography in (a) displaying the occurrence of the relative STM tip height. (d) Band structure of the Bi/Ag(111) surface alloy along the $\overline{\Gamma}\overline{K}$ measured by spin-resolved inverse photoemission. The data points indicate the peak positions with the solid lines as guides to the eye. Purple (blue) colors refer to the spin quantization axis forming a right (left) handed coordinate system with the $\overline{\Gamma}\overline{K}$-line, and the sample normal. (e) STM tunneling spectra (setpoint $V=-1\,$V, $I=1\,$nA, $V_{\rm mod}=25\,$mV) measured with the STM tip located at different surface locations, which are indicated in (a). The spectral features are labeled according to their band character.}
  \label{fig:2}
\end{figure*}

The \BiPbAg\ mixed surface alloy is a two-dimensional (2D) alloy of randomly mixed Bi and Pb atoms on the surface of Ag(111) \cite{bihlmayer2007enhanced}. The atomically resolved STM topography of the surface alloy is shown in Fig.\,\ref{fig:2}(a). It visualizes the random arrangement of Bi and Pb atoms on the triangular \sqthree\ alloy lattice, where they substitute every third Ag atom of the surface layer, leading to strong compositional disorder at any mixing ratio without clustering \cite{ast2008spin}. The Bi and Pb atoms retain their individual relaxation position outside the plane of the Ag(111) surface plane (Fig.\,\ref{fig:2}(b)) by which they can be distinguished in terms of their topographic height contrast $\Delta z$ in STM topography maps \cite{gierz2010structural}. Such analysis applied to the topography of Fig.\,\ref{fig:2}(a) is illustrated by the histogram in Fig.\,\ref{fig:2}(c) and reveals a \BixPbyAg{0.79}{0.21} mixing ratio of the alloy atoms.

The 2D electronic structure of this type of surface alloys is characterized by a set of three bands, $sp_z$, $m_j=1/2$, and $m_j=3/2$ \cite{bihlmayer2007enhanced}. We performed inverse-photoemission experiments on the Bi/Ag(111) (\BixPbyAg{x}{1-x}, $x=1$) surface alloy to characterize the band dispersion along the $\overline{\Gamma}\,\overline{\text{K}}$ direction (see Appendix\, \ref{appendix:methods}) \cite{wissing2014}. The measured dispersions of the unoccupied $m_j=1/2$ and $m_j=3/2$ bands shown in Fig.\,\ref{fig:2}(d) are consistent with results from {\em ab-initio} calculations for that platform \cite{bihlmayer2007enhanced}; please refer to Ref.\,\cite{ast2007giant} for the photo-emission data of the occupied $sp_z$-band in previous work. The band structure is visibly affected by strong Rashba spin-orbit coupling (SOC) \cite{ast2007giant}, and it is confined within the first atomic layer at the sample surface \cite{ast2007giant}, evidencing the 2D character of the surface alloy's electronic properties. 

Complete substitution of Bi atoms by Pb atoms in their lattice sites results in the formation of the pure Pb/Ag(111) surface alloy (\BixPbyAg{x}{1-x}, $x=0$). Owing to the different number of valence electrons in Bi ($6s^2$ $6p^3$) and Pb ($6s^2$ $6p^2$), the bands of a pristine ($x=0$) Pb/Ag(111) alloy ($sp_z$-onset at $E\approx650\,$meV) will shift by about 0.8\,eV to lower energy, when going to a pristine ($x=1$) Bi/Ag(111) alloy ($sp_z$-onset at $E\approx-135\,$meV) \cite{bihlmayer2007enhanced,ast2007local}. The partial substitution of Bi by Pb in the mixed surface alloy \BixPbyAg{0.79}{0.19}, therefore, results in a random disorder potential, whose spatial characteristics are determined by the position of the Pb atoms on the alloy lattice (Fig.\,\ref{fig:2}(a)). 

Scanning tunneling spectroscopy (STS) measurements can be used to characterize the influence of this random disorder potential on the LDOS. Fig.\ \ref{fig:2}(e) shows selected $dI/dV$-spectra, which were recorded in different regions of the mixed surface alloy surface, exhibiting a varying local Bi concentration (cf. Fig.\ \ref{fig:2}(a)). The spectra, all of which were recorded with the STM tip located on top of a Bi atom, exhibit similar spectral features that can be associated with the onsets of the respective electronic bands (cf.\ Fig.\ \ref{fig:2}(d)). However, we find the energetic position of these peaks to be shifted with respect to each other in the different tunneling spectra; their positions shift from higher to lower energy, when moving from regions of locally high Pb concentration to regions of locally high Bi concentration \cite{ast2007local}, consistent with the presence of a locally varying electrostatic potential. In passing we note that the tunneling spectra also do not exhibit a suppression of spectral weight around the Fermi level typically associated with electron-electron interactions in the presence of disorder \cite{efros1975coulomb, altshuler1980interaction, imry1982density, bielejec2001hard, richardella2010visualizing}.

To capture the real space characteristics of these disorder induced LDOS variations, we have performed spectroscopic imaging measurements with the STM, where we record $dI/dV$-maps over a larger surface area highlighted in Fig\,\ref{fig:3}(a) (see Appendix \,\ref{appendix:methods}). The normalized $dI/dV$-map measured at the $sp_z$-band onset at $E=80\,$meV, Fig.\,\ref{fig:3}(b), exhibits a highly irregular pattern with significant amplitude variations over nanometer length scales. We find that a high $dI/dV$ signal is predominantly accumulated in areas of high Pb concentration, consistent with the characteristics of the $dI/dV$-spectra in Fig.\,\ref{fig:2}(e). Conversely, the corresponding $dI/dV$-map measured deep inside the $sp_z$-band at $E=-400\,$meV, Fig.\,\ref{fig:3}(c), lacks a pronounced spatial structure, which is in agreement with the location-independent appearance of the individual $dI/dV$-spectra at this energy. We further find that the $dI/dV$-maps do not exhibit periodically ordered real space patterns, indicating the absence of charge density wave formation in the mixed surface alloy.

\begin{figure*}
  \includegraphics[width = 0.7\textwidth]{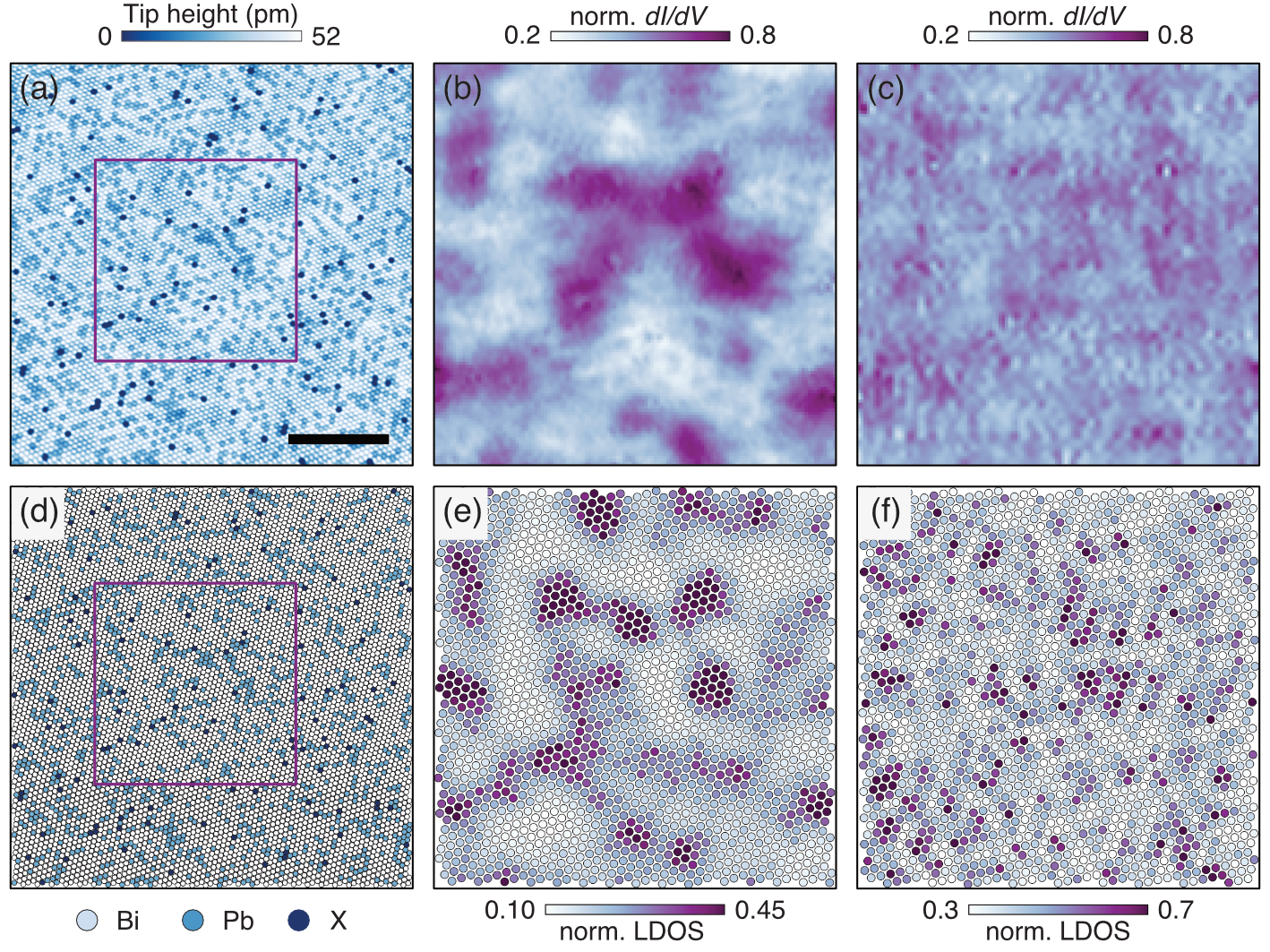}
  \caption{(a) STM topography of the \BixPbyAg{0.79}{0.21}\ mixed surface alloy (setpoint $V=3\,$mV, $I=7\,$nA). Scale bar 10\,nm. The purple box ($20\times20$\,nm$^2$) highlights the area in which the normalized tunneling conductance maps have been measured: (b) at the $sp_{\rm z}$-band edge ($+$0.08\,eV), (c) within the $sp_{\rm z}$-band ($-$0.4\,eV) (setpoint $I=1\,$nA, $V_{\rm mod}=25\,$mV). (d) Reconstructed lattice model of the mixed surface alloy where the topography in (a), serves as structural input for the Anderson tight-binding model calculation. Panel (e) and (f) show the calculated normalized LDOS maps at the $sp_{\rm z}$-band edge and inside the $sp_{\rm z}$-band, respectively (please refer to Table\,\ref{suptab:TBParams} for numerical parameters).}
  \label{fig:3}
\end{figure*}

Knowledge of disorder induced LDOS characteristics from model calculations can help to reveal the origin of the irregular $dI/dV$-pattern near the $sp_z$-band edge. We have constructed a realistic 2D Anderson tight binding model of the mixed surface alloy to calculate LDOS maps \cite{anderson1958}. Commonly, numerical investigations of the Anderson transition in two dimensions rely on artificial disorder distributions for the calculation of the eigenstates \cite{ando1989numerical, mildenberger2007wave}. In our study, we create the {\em realistic} Anderson model with a binary distribution of onsite disorder potentials, which is based on our detailed knowledge of the electronic properties and the alloy lattice structure from IPE and STM experiments. We obtain the nearest neighbor hopping potentials and Rashba terms from fits to the measured band dispersion in Fig.\,\ref{fig:2}(d). Moreover, our ability to distinguish between Bi and Pb atoms in STM measurements (c.f.\,Fig.\,\ref{fig:2}(c)) enables us to extract the exact atomic lattice structure of the topography in Fig.\,\ref{fig:3}(a). The reconstructed lattice is shown in Fig.\,\ref{fig:3}(d); we use it as a structural model input to solve for the wave functions $\Psi(\vec{r})$ and to calculate realistic LDOS $|\Psi(\vec{r})|^2$ maps (see Appendix\,\ref{appendix:tbm}). We also included a Rashba term, to account for the effect of strong SOC in the 2D mixed surface alloys. \cite{hikami1980spin, ast2007giant}. 

The calculated LDOS map near the $sp_z$-band edge is shown in Fig.\,\ref{fig:3}(e). It is in very good agreement with the corresponding $dI/dV$-map, Fig.\,\ref{fig:3}(b), down to the very detail. The calculated and measured maps deep inside the $sp_z$-band, Fig.\,\ref{fig:3}(c) and (f), exhibit a comparable amount of agreement despite their rather featureless appearance.

\section{Discussion}

The remarkable agreement between the $dI/dV$-map and LDOS-map shows that the measured $dI/dV$-signal predominantly reflects the spatial distribution of the 2D mixed surface alloy eigenstates density, i.e.\ the LDOS, with contributions coming from the Bi and Pb atoms only. This agreement is not least promoted by the absence of signatures of electronic correlations and CDW order in our experimental data in Fig.\,\ref{fig:2}(e) and Fig.\,\ref{fig:2}(b)-(c), respectively. While the absence of these phenomena could be explained by the elevated temperature of our experiment (100\,K), it is also consistent with the metallic character and the relatively simple topology of the $sp_{\rm z}$-band (c.f. Fig.\,\ref{fig:2}(d)) \cite{bihlmayer2007enhanced}.

Deviations between the measured and calculated LDOS could result from the simplicity of our Anderson model, which only considers isotropic nearest neighbor hopping, which is based on a single band, and which neglects any effect from the Ag substrate (see Appendix\,\ref{appendix:tbm}). The agreement between experiment and theory, therefore, emphasizes the predominant influence of the Pb atom induced random disorder potential on the spatial distribution of the mixed surface alloy's LDOS. This finding is also highlighted in the individual $dI/dV$-spectra in Fig.\,\ref{fig:2}(e), where the local Pb concentration affects the energetic position of the $sp_{\rm z}$-band edge. Consequently, these spectra miss the sharp Rashba-induced van-Hove singularity at the band onset, which was previously observed for both the pure Bi/Ag(111) and Pb/Ag(111) surface alloys \cite{ast2007local}. Importantly, we emphasize that the characteristic length scale of the measured and calculated LDOS near the band edge, Fig.\,\ref{fig:3}(b,e), is much larger than the atomic length scale of the disorder potential. The LDOS does not merely follow the local potential; instead, its structure emerges from the antagonistic interplay of localizing Pb impurities and delocalizing kinetic energy. 

Theoretically, for binary alloys, a disorder induced exponential tail in the average density of states is expected inside the (clean) band gap \cite{lifshitz1965energy}. This tail is formed by spatially localized impurity states, which (for the present case of a 2D system with spin-orbit coupling) are energetically separated by a sharp mobility edge from extended states far inside the band \cite{mott1968metal}. This picture is corroborated by our detailed analysis of the wave function decay characteristics based on the normalized partition ratio, NPR \cite{wegner1980inverse}: We find that the average wave function decay length is of atomic scale at the $sp_z$-band edge, and reaches the size of our measurement frame (20\,nm) at $E\approx -100\,$meV (see Appendix\,\ref{appendix:ipr}). The picture of a mobility edge also manifests in the spatial characteristics of the LDOS distributions; while the LDOS inside the localized impurity band and near the mobility edge exhibits a spatially irregular pattern, Fig.\,\ref{fig:3}(b) and (e), the LDOS deep inside the the $sp_{\rm z}$-band, Fig.\,\ref{fig:3}(c) and (f), appears more homogeneous, characteristic for the extended wave function of a metal.

\begin{figure*}
  \includegraphics[width = 0.7\textwidth]{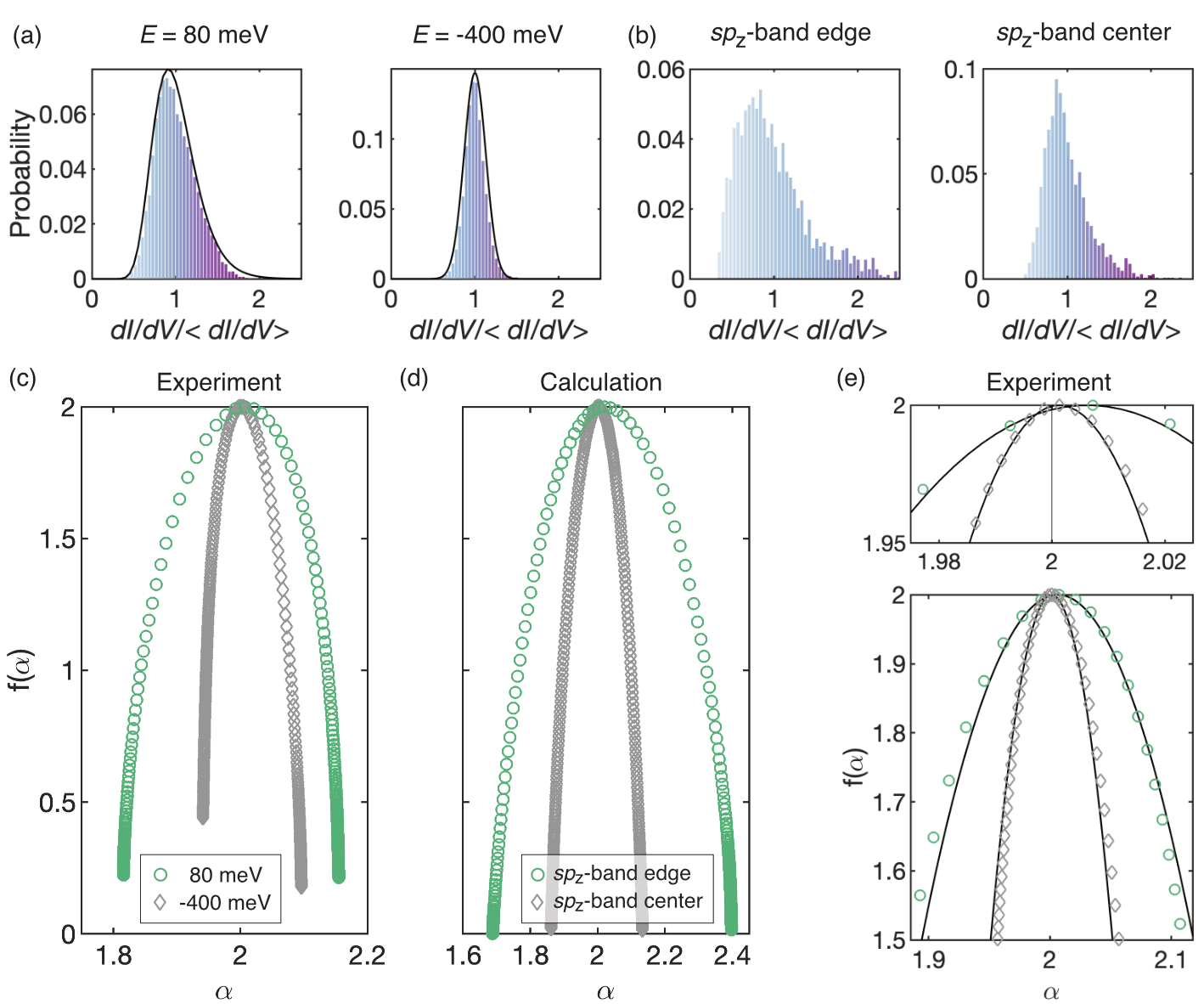}
  \caption{{\bf Multifractal analysis of the measured and calculated LDOS maps.} (a) Left panel: Statistical distribution of the tunnel conductance amplitude for the map measured at $V=80\,$ mV near the $sp_{\rm z}$-band edge in Fig.\,\ref{fig:3}(b). Right panel: The same analysis applied to the map measured at $V=-400\,$mV inside the $sp_{\rm z}$-band in Fig.\,\ref{fig:3}(c). The black lines represent the corresponding calculated distribution functions. (b) Left panel: Statistical distribution of the LDOS amplitude for the map calculated at the $sp_{\rm z}$-band edge in Fig.\,\ref{fig:3}(e). Right panel: The same analysis applied to the calculated map inside the $sp_{\rm z}$-band in Fig.\,\ref{fig:3}(f). (c) Multifractal singularity spectra $f(\alpha)$ of the tunneling conductance maps in Fig.\,\ref{fig:3}(b) (purple circles) and Fig.\,\ref{fig:3}(c) (blue diamonds). (d) Multifractal singularity spectra $f(\alpha)$ of the calculated LDOS maps in Fig.\,\ref{fig:3}(e) (purple circles) and Fig.\,\ref{fig:3}(f) (blue diamonds). (e) Bottom panel: Magnification of the $f(\alpha)$ spectra shown in (c). The black solid lines are the quadratic fits to the data. Top panel: Magnified view around the $f(\alpha)$ maxima.}
  \label{fig:4}
\end{figure*}

These observations can be further quantified by investigating the LDOS distribution function \cite{lerner1988distribution, mirlin1994distribution}. Whereas the LDOS of extended wavefunctions follows a Gaussian distribution, wavefunction multifractality yields a log-normal LDOS distribution, anticipating the broad distribution of the insulating state \cite{halsey1986fractal}. Applying such analysis to the measured LDOS maps of the $sp_z$ band, Fig.\,\ref{fig:3}(b) and (c), reveals their log-normal characteristics near the band edge, $\mu=\log(0.975)$ and $\sigma=0.25$), and, by contrast, their Gaussian characteristics, $\sigma=0.13$, inside the band (Fig.\,\ref{fig:4}(a)). We obtain similar characteristics for the LDOS distribution from model calculations, Fig.\,\ref{fig:4}(c). Even though both LDOS maps follow a log-normal distribution, their skewness is more pronounced near the $sp_{\rm z}$-band edge as compared to inside the $sp_{\rm z}$-band, Fig.\,\ref{fig:4}(c).

The corresponding multifractal singularity spectrum, $f(\alpha)$, can be directly calculated from the LDOS maps as a function of the singularity strength $\alpha$, see Appendix\,\ref{appendix:mfa} \cite{chhabra1989direct}. In Fig.\,\ref{fig:4}(c) we plot $f(\alpha)$ of the measured LDOS maps. The broadened parabolic appearance of $f(\alpha)$, centered at $\alpha>2$, supports that the mixed surface alloy LDOS exhibits multifractal scaling behavior. $f(\alpha)$ of the LDOS map near the $sp_z$-band edge, Fig.\,\ref{fig:4}(c), is significantly wider than the $f(\alpha)$ of the LDOS map measured inside the $sp_z$-band. This observation is consistent with the different character of the LDOS distribution functions in Fig.\,\ref{fig:4}(a) and with the highly irregular (homogeneous) real space pattern of the LDOS maps in Fig.\,\ref{fig:3}(b) (Fig.\,\ref{fig:3}(c)). $f(\alpha)$ of the calculated LDOS maps, Fig.\,\ref{fig:4}(d), exhibit a wider shape than their experimental counterparts, consistent with the corresponding deviations between the LDOS distributions, Fig.\,\ref{fig:4}(a) and (b), but show overall very good qualitative agreement. The deviations could, on the one hand, be rooted in the discussed simplifications of the Anderson model. On the other hand, the finite temperature of our experiment (100\,K) promotes electron-phonon scattering induced quantum decoherence, weakening the eigenstate multifractality in the measured LDOS maps.

The anomalous eigenstate scaling distinguishes the quantum critical wave functions from those in the metallic phase at $\alpha=2$ \cite{evers2008anderson}. It has been theoretically shown that the anomalous scaling of the Wigner-Dyson classes (WDC) exhibits an exact symmetry \cite{mirlin2006exact}, which, for a 2D system, manifests in the relation, $f(\alpha)=f(4-\alpha)-2+\alpha$.
At weak multifractality, the associated singularity spectrum, $f(\alpha)=2-(\alpha-2-\gamma)^2/(4\gamma)$, is parabolic, where $\gamma$ is a measure for the strength of multifractality and defines both the maximum at $\alpha=2+\gamma$ and the opening angle of the parabola \cite{evers2008anderson}. We note that the mixed surface alloy, which preserves time-reversal symmetry and whose electronic states are strongly influenced by SOC \cite{bihlmayer2007enhanced, ast2007giant}, falls into the symplectic WDC AII for which numerical calculations predict $\gamma = 0.172$ \cite{evers2008anderson}. 

Our quantitative analysis of the experimental $f(\alpha)$ in Fig.\,\ref{fig:4}(e) reveals their predominently parabolic characteristics and consistency with the theoretical expectation of an exact anomalous scaling symmetry. Applying a quadratic fit to $f(\alpha)$ we obtain $\gamma_\text{exp}=(6.3\pm0.4)\times10^{-3}$ ($\gamma_\text{calc}=(21.6\pm1.1)\times10^{-3}$) and $\gamma_\text{exp}=(1.30\pm0.05)\times10^{-3}$ ($\gamma_\text{calc}=(2.60\pm0.05)\times10^{-3}$) for the experimental (calculated) LDOS map near the edge and inside the $sp_{\rm z}$-band. We note that non-parabolic corrections for $f(\alpha)<1.5$, Fig.\,\ref{fig:4}(c) and (d), are in agreement with previous theoretical work on the scaling symmetry \cite{mildenberger2007wave}. 

The extracted small values of $\gamma$, on the one hand, reflect moderate multifractality below the theoretically expected extent. It is likely that the previously discussed finite experimental temperature (100\,K) weakens the multifractal eigenstate characteristics. What is more, our measurement also integrates the eigenstates over a 70\,meV window, including those states deeper inside the band, which exhibit extended state characteristics with reduced multifractal character, cf. normalized partition ratio analysis in Appendix\,\ref{appendix:ipr}. In particular, the normalized disorder strength in the mixed surface alloy $W/t=0.7$ ($W$ is the difference between Bi and Pb onsite potentials, $t$ is the nearest neighbor hopping) such that most eigenstates in the mixed surface alloy $sp_{z}$-band are delocalized and away from the critical point. By comparison, highly accurate model calculations {\em at criticality} with $W/t>5$ show an enhancement of $\gamma$ by almost a factor of 10 that manifests in an accordingly larger degree of wavefunction multifractality \cite{ando1989numerical, mildenberger2007wave}. 

Finally, in order to estimate the impact of the mentioned experimental limitations, we analyze the $dI/dV$-map measured at $E=1\,$eV near the $m_{\rm j}=3/2$-band onset. Here one would also expect multifractal wavefunction characteristics similar to those observed near the $sp_{\rm z}$-band edge, Fig.\,\ref{fig:3}(b) and Fig.\,\ref{fig:4}(a), but these characteristics are not detected, Appendix\,\ref{appendix:mj32}. Even though the real space pattern of the $dI/dV$-map, Fig.\,\ref{fig:7}(a), appears irregular, areas of high and low $dI/dV$-amplitude, respectively are found with similar occurrence; this is also reflected in the quasi-Gaussian distribution in Fig.\,\ref{fig:7}(b). This discrepency of observations can be understood in terms of Fermi liquid theory, according to which quasiparticle scattering is strongly enhanced at 1eV away from the Fermi energy, which promotes eigenstate decoherence, detrimental to the observation of MF characteristics. \textit{Argumentum a contrario}, we take this discussion as evidence for sufficient quantum coherence for the detection of multifractality at the $sp_{\rm z}$ edge.

\section{Conclusion}
We combined STM and IPE experiments with Anderson model calculations to study the real space LDOS characteristics of the 2D mixed surface alloy \BixPbyAg{0.79}{0.21}. Spectroscopic measurements of the LDOS with the STM reveal a strong influence of a locally varying disorder potential on the electronic states that is induced by the different valency of the Bi and Pb atoms.

The effect of disorder-induced wavefunction localization in the mixed surface alloy is quantified by the statistical analysis of the spatial LDOS distributions . We observe the log-normal characteristics of localized states near the $sp_{\rm z}$-band edge, and the Gaussian characteristics of metallic states inside the $sp_{\rm z}$-band, indicating the presence of a mobility edge. The singularity spectra further reveal multifractal scaling of the eigenstate amplitude, which we find to be significant near the $sp_{\rm z}$-band edge. 

We can reproduce these observations with very good agreement through realistic Anderson model calculations, which are based on our detailed knowledge of the electronic properties and the atomic scale lattice structure of the surface alloy. The same analysis further confirms the symmetry of the anomalous eigenstate scaling at criticality in the Wigner-Dyson classes predicted by theory.

Looking ahead, realizing a larger disorder strength in this material system--for instance by moving beyond a binary alloy and by combining suitable elements with enhanced on-site potential differences--would be desirable, because such systems could provide deeper insights into strong multifractal eigenstate scaling \cite{evers2008anderson, mildenberger2007wave}. We anticipate that performing these and related experiments at much reduced temperatures should suppress scattering-induced eigenstate decoherence and enhance the detection of multifractal characteristics. Recent experiments on 2D films of disordered transition-metal dichalcogenides also suggest the existence of multifractal superconductivity \cite{zhao2019disorder, rubio2018visualization}. Applying our methodology to these systems could provide unique insights on the underlying normal state eigenstate multifractality electronic states and the correlation with the superconducting critical temperature \cite{feigel2007eigenfunction, feigel2010fractal, burmistrov2012enhancement, mayoh2015global}. In this regard, the actively studied material class of the superconducting high entropy alloys could represent avenues to explore the interplay of wavefunction multifractality at strong compositional disorder with the properties of the superconducting order parameter \cite{kovzelj2014discovery, sun2019high}.

\begin{acknowledgements}
We gratefully acknowledge stimulating discussions with Maurits W.\ Haverkort, M.\ Assig, C.\ Stra\ss er, M.\ Etzkorn, H.\ Benia, M. Eltschka, and A.\ Schnyder. B.\ J.\ acknowledges funding from the Alexander-von-Humboldt foundation during the completion of this work. E.\ J.\ K.\ acknowledges support by DOE Basic Energy Sciences grant DE-FG02-99ER45790. C.\ R.\ A.\ acknowledges funding from the Emmy-Noether-Program of the Deutsche Forschungsgemeinschaft (DFG).
\end{acknowledgements}

\appendix
\section{Materials and Methods}
\label{appendix:methods}
Preparation of the \BiPbAg\ mixed surface alloys was done by the co-evaporation of sub-monolayer amounts of Bi and Pb atoms on an atomically clean Ag(111) surface. The Ag substrate was cleaned by subsequent cycles of argon sputtering and thermal annealing. During the alloy deposition, the sample was kept at a temperature of 573\,K for optimal mobility of the atoms to form a long range ordered alloy. Sample preparation and measurements were done in the ultra high vacuum environment, $p\geq10^{-10}\,$mbar, of a home-built scanning tunneling microscope. The temperature of tip and sample during measurements was 100\,K. The tunnel conductance spectra and maps were measured at appropriate bias $V$ and current $I$ setpoints, respectively indicated in the figure captions, using a lock-in modulation amplitude, $V_{\rm mod}=25\,$mV. A flat background was subtracted from the tunnel conductance map to account for the finite and spatially homogeneous background LDOS from the Ag(111) bulk states. The overall energy resolution of our experiment was determined to be about 70\,meV. 

Spin-resolved IPE was performed at room temperature using spin-polarized electrons emitted from a GaAs photocathode \cite{budke2007inverse}. The photons were detected with a Geiger-M\"uller counter at $E=9.8\,$eV \cite{thiede2018acetone}. The overall IPE energy resolution was 350\,meV and the angular resolution $\pm2^{\circ}$ \cite{zumbulte2015momentum}.

\section{Anderson tight binding model}
\label{appendix:tbm}
We calculate theoretical LDOS maps using a realistic Anderson tight binding model of the 2D mixed surface alloy \cite{alvermann2005local, schubert2010distribution,anderson1958}. We use a single $p_z$-orbital model with on-site potential $\epsilon_{\rm i}$ and a nearest neighbor hopping potential $t$, which is isotropic on the alloy's two-dimensional hexagonal lattice. To simplify the model calculation, we also assume identical hopping potentials for all different nearest neighbor pairs and we neglect coupling of the alloy atoms to the Ag substrate. Strong SOC in the \BiPbAg\ mixed surface alloy \cite{ast2007giant} is considered by  a nearest-neighbor Rashba-type spin-splitting with coupling constant $\alpha_R$ \cite{hikami1980spin, asada2002anderson}:

\begin{widetext}
\begin{equation}
H=\sum\limits_{i,\sigma_i} \epsilon_i\ |i,\sigma_i\rangle\langle i,\sigma_i| + \sum\limits_{\langle i,j\rangle,\sigma_i=\sigma_j} t\ |i,\sigma_i\rangle\langle j,\sigma_j|
+\sum\limits_{\langle i,j\rangle,\sigma_i\neq\sigma_j} \alpha_R\left(\vec{\sigma}\times\vec{\tau}_{ij}\right)_z |i,\sigma_i\rangle\langle j,\sigma_j|
\label{eq:tbham}
\end{equation}
\end{widetext}

The angle parentheses in the second sum limits indicate summation over nearest neighbors, $\vec{\sigma}$ is the vector of Pauli matrices, $\vec{\tau}_{ij}$ is a unit vector connecting two neighboring atoms, and $\sigma_i$ is the spin index. We use the reconstructed atomic lattice structure in Fig.\,\ref{fig:3}(d) as structural model input to assign the Bi and Pb specific on-site potentials $\epsilon_{\rm i}$. The tight-binding calculation gives access to the complete set of eigenstates $\Psi(\vec{r}, E)$ of the mixed surface alloy that can be used to calculate spatially resolved eigenstate density from the eigenstate spinors,
\begin{equation}
|\Psi(\vec{r},E)|^2=|\Psi(\vec{r},E,\uparrow)|^2+|\Psi(\vec{r},E,\downarrow)|^2.
\end{equation}
Realistic LDOS maps (cf.\ Fig.\ \ref{fig:3}(e,f)) are obtained by convolving the spatially resolved eigenstate densities by a Gaussian function ($\sigma=35\,$meV) to account for the finite temperature and energy resolution of our experiment.

\begin{figure}
  \includegraphics[width = .9\columnwidth]{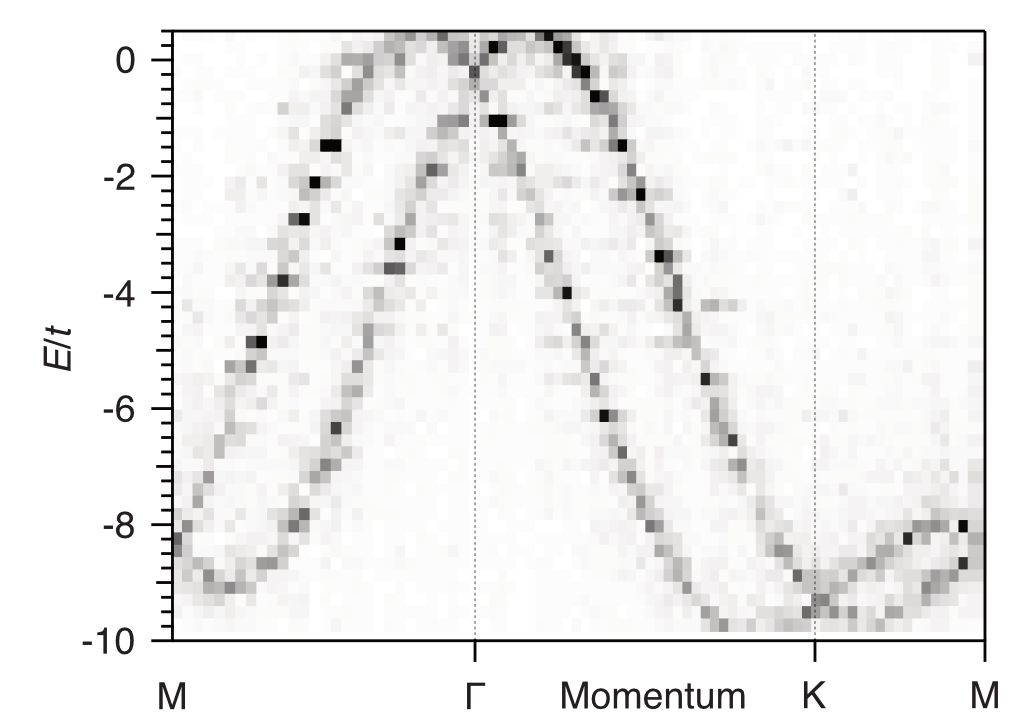}
  \caption{Calculated band structure of the $sp_{\rm z}$-band. The corresponding tight-binding parameter are given in Table\,\ref{suptab:TBParams}. The horizontal lines mark the energies at which the LDOS maps were calculated.}
  \label{fig:5}
\end{figure}

We obtain realistic values for the Rashba and hopping potentials by fitting the calculated surface alloy bands to the experimentally determined band dispersions (Fig.\,\ref{fig:1}(d)). The band dispersion in momentum space can be calculated by a Fourier transformation of the eigenstates obtained from the Anderson tight-binding model under periodic boundary conditions \cite{ ku2010unfolding,popescu2010effective, haverkort2011electronic}. The calculated $sp_{\rm z}$-band dispersion, which is shown in Fig.\,\ref{fig:5}, is consistent with results from {\em ab-initio} calculations \cite{bihlmayer2007enhanced} and from photo-emission experiments \cite{ast2007giant}. Finally, we obtain a realistic estimate for the Bi and Pb on-site potential difference $\Delta\epsilon=\epsilon_{\rm Pb}-\epsilon_{\rm Bi}$ from the relative energy shift of the band onsets in the tunneling conductance spectra in Fig.\,\ref{fig:2}(e). The complete parameter set for the Anderson tight binding model is given in Table \ref{suptab:TBParams}.

\begin{table}
    \begin{tabular}{c@{\extracolsep{2mm}}rrrrrrrr}
      band & $E_{\text{exp}}$ & $\epsilon_{\text{Bi}}$ & $\epsilon_{\text{Pb}}$ & $\epsilon_{\text{def}}$ & $t$ & $\alpha_R$ & $E_{\text{pos}}$ & $E_{\text{broad}}$ \\
      \hline
      \hline
      $sp_z$    & 0.08    & $-$0.14 & 0.14 & 0.32 & 0.4    & 0.3 & $-$0.06 & 0.07 \\
      $sp_z$    & $-$0.40 & $-$0.14 & 0.14 & 0.32 & 0.4    & 0.3 & $-$0.6 & 0.07 \\
      \hline
      \hline
    \end{tabular}
    \caption{Parameters used in the tight-binding calculation for the different local density of states (LDOS) maps. The first column indicates the band character. $E_{\text{exp}}$ is the energy at which the experimental LDOS map has been measured,  $\epsilon_{\text{Bi}}$ and $\epsilon_{\text{Pb}}$ are the onsite potentials for the Bi and Pb atoms, respectively.  $\epsilon_{\text{def}}$ is the onsite potential of an defect site, where no alloy atom is substituted (black dots in the topography). $t$ is the hopping parameter, $\alpha_R$ is the Rashba parameter, $E_{\text{pos}}$ is the energy shift from the band edge at which the LDOS map has been calculated, and $E_{\text{broad}}$ is the full width at half maximum value for the Gaussian broadening with which the calculated data has been convoluted. All values are given in electronvolts.}
    \label{suptab:TBParams}
\end{table}

\section{Multifractal analysis}
\label{appendix:mfa}

The multifractal singularity spectrum $f(\alpha)$ can be  calculated from 2D maps using a method developed by Chhabra and Jensen \cite{chhabra1989direct}. This analysis is based on the numerical evaluation of the normalized local box probabilities $\mu_{\rm j}(\lambda, q)$,
\begin{equation}
    \mu_{j}(\lambda, q)=\frac{1}{\sum_{k=1}^{N(\lambda)}P_{k}(\lambda)^q}P_{ j}(\lambda)^q,
\end{equation}
where $P_{\rm k}(\lambda)^q$ denotes the box probability of box $k$ for the distribution LDOS moment $q$. The square shaped box with edge length, $L_{\rm k}$, is characterized by its normalized size, $\lambda=L_{\rm k}/L_{\rm m}$, determining the amount of boxes, $N$, inside a given square shaped LDOS map of edge length $L_{\rm m}$. Put into the context of the LDOS maps, $P_{\rm k}(\lambda)^q$ denotes the sum of the LDOS spectral weight inside a box of size $\lambda$ at position $k$.

Knowledge of $\mu_{\rm j}(\lambda, q)$ enables the direct calculation of $f(\alpha)$,
\begin{equation}
    f(\alpha(q))=\lim_{\lambda \to 0}\frac{\sum_{j=1}^{N(\lambda)}\mu_{j}(\lambda, q)\ln{(\mu_{j}(\lambda, q))}}{\ln{(\lambda)}},
\end{equation}
and the singularity strength, $\alpha(q)$,
\begin{equation}
    \alpha(q)=\lim_{\lambda \to 0}\frac{\sum_{j=1}^{N(\lambda)}\mu_{j}(\lambda, q)\ln{(P_{\rm j}(\lambda)^q)}}{\ln{(\lambda)}}.
\end{equation}
We used $\lambda=0.1$ and a moment interval $q=[-200, 200]$ for the calculation of the $f(\alpha)$ shown in Fig.\,\ref{fig:4}(c,d).

\section{Normalized partition ratio analysis}
\label{appendix:ipr}

\begin{figure}[t]
  \includegraphics[width = 1\columnwidth]{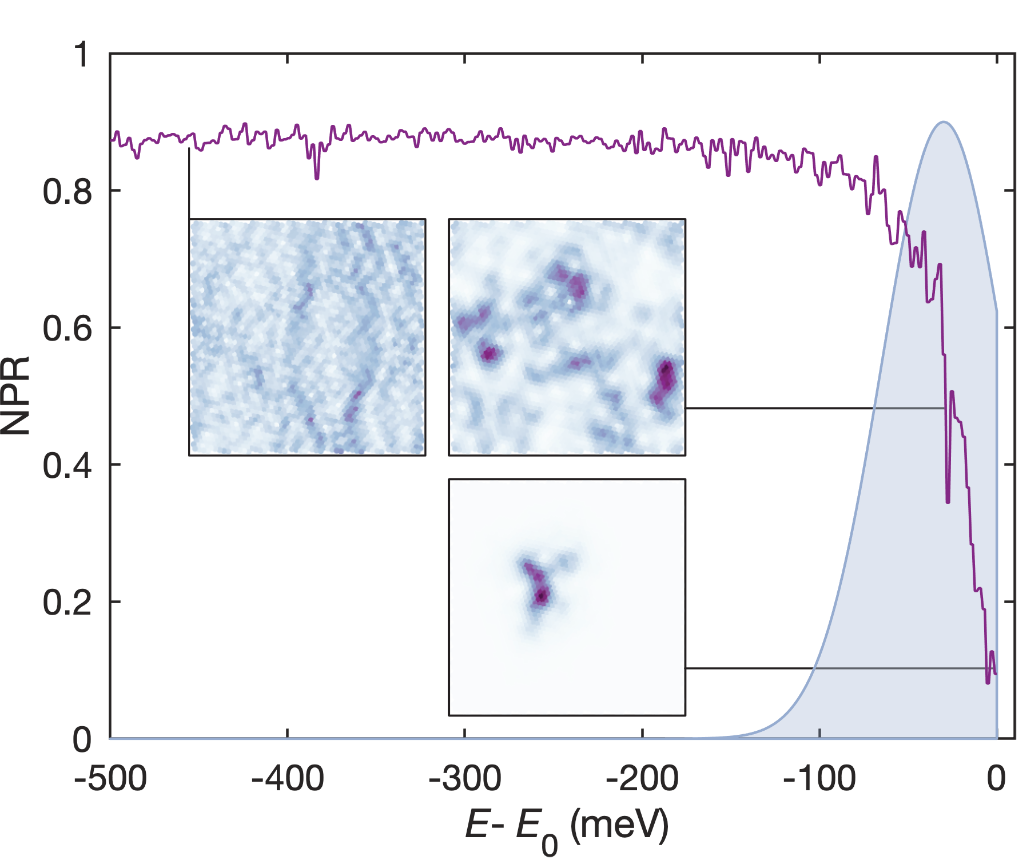}
  \caption{Normalized partition ratio (NPR) as a function of the energy, $E$, from the band onset, $E_0=0$, for each calculated eigenstate in the $sp_{\rm z}$-band. The shaded Gaussian shows the weighted integration interval for the calculation of the LDOS map in Fig.\ref{fig:3}(e)). The insets show the spatially resolved eigenstate densities, $|\Psi(\vec{r}, E)|^2$, at the respective energies (purple/white: high/low values, edge length 20\,nm).}
  \label{fig:6}
\end{figure}

The normalized partition ratio (NPR) can be used as a measure to quantify eigenstate localization in a given physical system. For a given eigenstate of a tight-binding-model, it is defined as
\begin{equation}
{\rm NPR}=1/(N\sum\limits_{\vec{r}}|\Psi(\vec{r})|^{4}),
\end{equation}
where $N$ denotes the number of lattice sites and $\sum\limits_{\vec{r}}|\Psi(\vec{r})|^2=1$ \cite{wegner1980inverse}. The NPR describes the fraction of system sites, which contribute to the spectral weight of an eigenstate. For a completely extended state, which exhibits plane wave characteristics, ${\rm NPR}\to1$.The more localized the state is, the smaller the NPR becomes. For a fully localized state, ${\rm NPR}\to0$. Fig.\,\ref{fig:6} shows the NPR, which was calculated for the Anderson tight-binding model eigenstates of the $sp_{\rm z}$-band as a function of the energy separation from the band onset at $E_{\rm 0}=0\,$meV. We also plot the spatially resolved density distribution, $|\Psi(\vec{r})|^2$, of selected eigenstates in the insets. ${\rm NPR}\approx0.1$ for the eigenstates near the band onset signals eigenstate localization, consistent with the spatial characteristics of the corresponding $|\Psi(\vec{r})|^2$. Moving inside the band, the NPR increases, until it plateaus at ${\rm NPR}\approx0.9$ deep inside the $sp_{\rm z}$-band, a trend which is also reflected in the real space characteristics of the $|\Psi(\vec{r})|^2$.

\begin{figure}
  \includegraphics[width = 1\columnwidth]{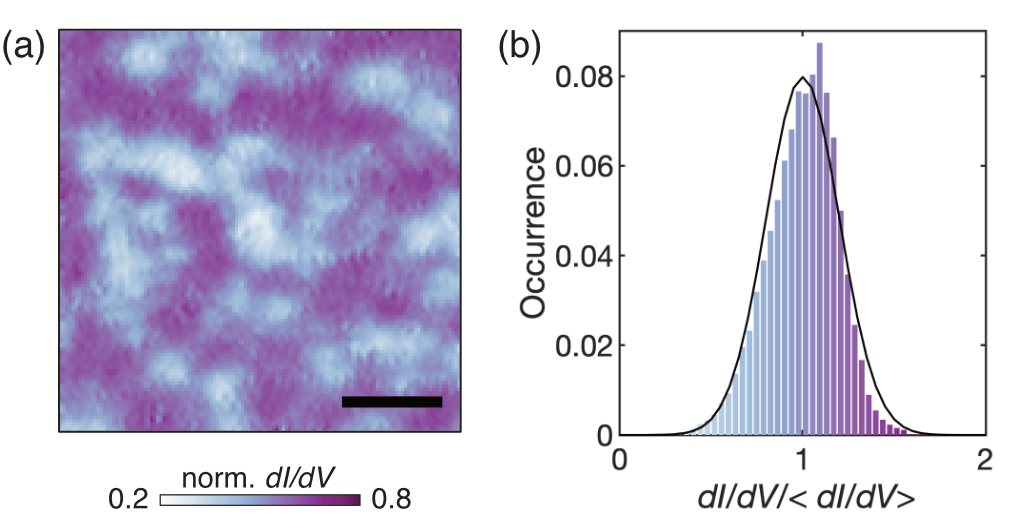}
  \caption{(a) Normalized tunnel conductance map (area: purple box in STM topography of Fig.\,\ref{fig:3}(a)) measured around the band minimum of the $m_{\rm j}=3/2$-band (setpoint $V=1\,$V, $I=1\,$nA, $V_{\rm mod}=25\,$mV), scale bar 5\,nm). (b) Statistical distribution of the corresponding tunnel conductance amplitude.}
  \label{fig:7}
\end{figure}

\section{Quasiparticle lifetime effects.}
\label{appendix:mj32}

To illustrate the effect of quasiparticle scattering on the real space characteristics of the LDOS away from Fermi energy, we have recorded a tunneling conductance map near the $m_{\rm j}=3/2$-band edge at $E=1\,$eV shown in Fig\,\ref{fig:7}(a)). The corresponding LDOS distribution histogram is shown in Fig\,\ref{fig:7}(b)). It is slightly asymmetric, but it can essentially be described by a Gaussian distribution, indicating the absence of multifractal scaling of the eigenstate real space distribution.

\bibliography{refs} 

\begin{thebibliography}{51}%
\makeatletter
\providecommand \@ifxundefined [1]{%
 \@ifx{#1\undefined}
}%
\providecommand \@ifnum [1]{%
 \ifnum #1\expandafter \@firstoftwo
 \else \expandafter \@secondoftwo
 \fi
}%
\providecommand \@ifx [1]{%
 \ifx #1\expandafter \@firstoftwo
 \else \expandafter \@secondoftwo
 \fi
}%
\providecommand \natexlab [1]{#1}%
\providecommand \enquote  [1]{``#1''}%
\providecommand \bibnamefont  [1]{#1}%
\providecommand \bibfnamefont [1]{#1}%
\providecommand \citenamefont [1]{#1}%
\providecommand \href@noop [0]{\@secondoftwo}%
\providecommand \href [0]{\begingroup \@sanitize@url \@href}%
\providecommand \@href[1]{\@@startlink{#1}\@@href}%
\providecommand \@@href[1]{\endgroup#1\@@endlink}%
\providecommand \@sanitize@url [0]{\catcode `\\12\catcode `\$12\catcode
  `\&12\catcode `\#12\catcode `\^12\catcode `\_12\catcode `\%12\relax}%
\providecommand \@@startlink[1]{}%
\providecommand \@@endlink[0]{}%
\providecommand \url  [0]{\begingroup\@sanitize@url \@url }%
\providecommand \@url [1]{\endgroup\@href {#1}{\urlprefix }}%
\providecommand \urlprefix  [0]{URL }%
\providecommand \Eprint [0]{\href }%
\providecommand \doibase [0]{http://dx.doi.org/}%
\providecommand \selectlanguage [0]{\@gobble}%
\providecommand \bibinfo  [0]{\@secondoftwo}%
\providecommand \bibfield  [0]{\@secondoftwo}%
\providecommand \translation [1]{[#1]}%
\providecommand \BibitemOpen [0]{}%
\providecommand \bibitemStop [0]{}%
\providecommand \bibitemNoStop [0]{.\EOS\space}%
\providecommand \EOS [0]{\spacefactor3000\relax}%
\providecommand \BibitemShut  [1]{\csname bibitem#1\endcsname}%
\let\auto@bib@innerbib\@empty
\bibitem [{\citenamefont {Sreenivasan}(1991)}]{sreenivasan1991}%
  \BibitemOpen
  \bibfield  {author} {\bibinfo {author} {\bibfnamefont {K\_R}\ \bibnamefont
  {Sreenivasan}},\ }\bibfield  {title} {\enquote {\bibinfo {title} {Fractals
  and multifractals in fluid turbulence},}\ }\href@noop {} {\bibfield
  {journal} {\bibinfo  {journal} {Annual Review of Fluid Mechanics}\ }\textbf
  {\bibinfo {volume} {23}},\ \bibinfo {pages} {539--604} (\bibinfo {year}
  {1991})}\BibitemShut {NoStop}%
\bibitem [{\citenamefont {Stanley}\ and\ \citenamefont
  {Meakin}(1988)}]{stanley1988multifractal}%
  \BibitemOpen
  \bibfield  {author} {\bibinfo {author} {\bibfnamefont {H~Eugene}\
  \bibnamefont {Stanley}}\ and\ \bibinfo {author} {\bibfnamefont {Paul}\
  \bibnamefont {Meakin}},\ }\bibfield  {title} {\enquote {\bibinfo {title}
  {Multifractal phenomena in physics and chemistry},}\ }\href@noop {}
  {\bibfield  {journal} {\bibinfo  {journal} {Nature}\ }\textbf {\bibinfo
  {volume} {335}},\ \bibinfo {pages} {405--409} (\bibinfo {year}
  {1988})}\BibitemShut {NoStop}%
\bibitem [{\citenamefont {Abrahams}\ \emph {et~al.}(1979)\citenamefont
  {Abrahams}, \citenamefont {Anderson}, \citenamefont {Licciardello},\ and\
  \citenamefont {Ramakrishnan}}]{abrahams1979scaling}%
  \BibitemOpen
  \bibfield  {author} {\bibinfo {author} {\bibfnamefont {Elihu}\ \bibnamefont
  {Abrahams}}, \bibinfo {author} {\bibfnamefont {PW}~\bibnamefont {Anderson}},
  \bibinfo {author} {\bibfnamefont {DC}~\bibnamefont {Licciardello}}, \ and\
  \bibinfo {author} {\bibfnamefont {TV}~\bibnamefont {Ramakrishnan}},\
  }\bibfield  {title} {\enquote {\bibinfo {title} {Scaling theory of
  localization: Absence of quantum diffusion in two dimensions},}\ }\href@noop
  {} {\bibfield  {journal} {\bibinfo  {journal} {Physical Review Letters}\
  }\textbf {\bibinfo {volume} {42}},\ \bibinfo {pages} {673} (\bibinfo {year}
  {1979})}\BibitemShut {NoStop}%
\bibitem [{\citenamefont {Wegner}(1979)}]{wegner1979mobility}%
  \BibitemOpen
  \bibfield  {author} {\bibinfo {author} {\bibfnamefont {Franz}\ \bibnamefont
  {Wegner}},\ }\bibfield  {title} {\enquote {\bibinfo {title} {The mobility
  edge problem: continuous symmetry and a conjecture},}\ }\href@noop {}
  {\bibfield  {journal} {\bibinfo  {journal} {Zeitschrift f{\"u}r Physik B
  Condensed Matter}\ }\textbf {\bibinfo {volume} {35}},\ \bibinfo {pages}
  {207--210} (\bibinfo {year} {1979})}\BibitemShut {NoStop}%
\bibitem [{\citenamefont {Anderson}(1958)}]{anderson1958}%
  \BibitemOpen
  \bibfield  {author} {\bibinfo {author} {\bibfnamefont {Philip~W}\
  \bibnamefont {Anderson}},\ }\bibfield  {title} {\enquote {\bibinfo {title}
  {Absence of diffusion in certain random lattices},}\ }\href@noop {}
  {\bibfield  {journal} {\bibinfo  {journal} {Physical Review}\ }\textbf
  {\bibinfo {volume} {109}},\ \bibinfo {pages} {1492} (\bibinfo {year}
  {1958})}\BibitemShut {NoStop}%
\bibitem [{\citenamefont {Castellani}\ and\ \citenamefont
  {Peliti}(1986)}]{castellani1986multifractal}%
  \BibitemOpen
  \bibfield  {author} {\bibinfo {author} {\bibfnamefont {C}~\bibnamefont
  {Castellani}}\ and\ \bibinfo {author} {\bibfnamefont {L}~\bibnamefont
  {Peliti}},\ }\bibfield  {title} {\enquote {\bibinfo {title} {Multifractal
  wavefunction at the localisation threshold},}\ }\href@noop {} {\bibfield
  {journal} {\bibinfo  {journal} {Journal of Physics A: Mathematical and
  General}\ }\textbf {\bibinfo {volume} {19}},\ \bibinfo {pages} {L429}
  (\bibinfo {year} {1986})}\BibitemShut {NoStop}%
\bibitem [{\citenamefont {Lerner}(1988)}]{lerner1988distribution}%
  \BibitemOpen
  \bibfield  {author} {\bibinfo {author} {\bibfnamefont {IV}~\bibnamefont
  {Lerner}},\ }\bibfield  {title} {\enquote {\bibinfo {title} {Distribution
  functions of current density and local density of states in disordered
  quantum conductors},}\ }\href@noop {} {\bibfield  {journal} {\bibinfo
  {journal} {Physics Letters A}\ }\textbf {\bibinfo {volume} {133}},\ \bibinfo
  {pages} {253--259} (\bibinfo {year} {1988})}\BibitemShut {NoStop}%
\bibitem [{\citenamefont {Mirlin}\ and\ \citenamefont
  {Fyodorov}(1994)}]{mirlin1994distribution}%
  \BibitemOpen
  \bibfield  {author} {\bibinfo {author} {\bibfnamefont {Alexander~D}\
  \bibnamefont {Mirlin}}\ and\ \bibinfo {author} {\bibfnamefont {Yan~V}\
  \bibnamefont {Fyodorov}},\ }\bibfield  {title} {\enquote {\bibinfo {title}
  {Distribution of local densities of states, order parameter function, and
  critical behavior near the anderson transition},}\ }\href@noop {} {\bibfield
  {journal} {\bibinfo  {journal} {Physical Review Letters}\ }\textbf {\bibinfo
  {volume} {72}},\ \bibinfo {pages} {526} (\bibinfo {year} {1994})}\BibitemShut
  {NoStop}%
\bibitem [{\citenamefont {Halsey}\ \emph {et~al.}(1986)\citenamefont {Halsey},
  \citenamefont {Jensen}, \citenamefont {Kadanoff}, \citenamefont {Procaccia},\
  and\ \citenamefont {Shraiman}}]{halsey1986fractal}%
  \BibitemOpen
  \bibfield  {author} {\bibinfo {author} {\bibfnamefont {Thomas~C}\
  \bibnamefont {Halsey}}, \bibinfo {author} {\bibfnamefont {Mogens~H}\
  \bibnamefont {Jensen}}, \bibinfo {author} {\bibfnamefont {Leo~P}\
  \bibnamefont {Kadanoff}}, \bibinfo {author} {\bibfnamefont {Itamar}\
  \bibnamefont {Procaccia}}, \ and\ \bibinfo {author} {\bibfnamefont {Boris~I}\
  \bibnamefont {Shraiman}},\ }\bibfield  {title} {\enquote {\bibinfo {title}
  {Fractal measures and their singularities: The characterization of strange
  sets},}\ }\href@noop {} {\bibfield  {journal} {\bibinfo  {journal} {Physical
  Review A}\ }\textbf {\bibinfo {volume} {33}},\ \bibinfo {pages} {1141}
  (\bibinfo {year} {1986})}\BibitemShut {NoStop}%
\bibitem [{\citenamefont {Evers}\ and\ \citenamefont
  {Mirlin}(2008)}]{evers2008anderson}%
  \BibitemOpen
  \bibfield  {author} {\bibinfo {author} {\bibfnamefont {Ferdinand}\
  \bibnamefont {Evers}}\ and\ \bibinfo {author} {\bibfnamefont {Alexander~D}\
  \bibnamefont {Mirlin}},\ }\bibfield  {title} {\enquote {\bibinfo {title}
  {Anderson transitions},}\ }\href@noop {} {\bibfield  {journal} {\bibinfo
  {journal} {Reviews of Modern Physics}\ }\textbf {\bibinfo {volume} {80}},\
  \bibinfo {pages} {1355} (\bibinfo {year} {2008})}\BibitemShut {NoStop}%
\bibitem [{\citenamefont {Mandelbrot}(1974)}]{mandelbrot1974intermittent}%
  \BibitemOpen
  \bibfield  {author} {\bibinfo {author} {\bibfnamefont {Benoit~B}\
  \bibnamefont {Mandelbrot}},\ }\bibfield  {title} {\enquote {\bibinfo {title}
  {Intermittent turbulence in self-similar cascades: divergence of high moments
  and dimension of the carrier},}\ }\href@noop {} {\bibfield  {journal}
  {\bibinfo  {journal} {Journal of Fluid Mechanics}\ }\textbf {\bibinfo
  {volume} {62}},\ \bibinfo {pages} {331--358} (\bibinfo {year}
  {1974})}\BibitemShut {NoStop}%
\bibitem [{\citenamefont {Hentschel}\ and\ \citenamefont
  {Procaccia}(1983)}]{hentschel1983infinite}%
  \BibitemOpen
  \bibfield  {author} {\bibinfo {author} {\bibfnamefont {HGE}\ \bibnamefont
  {Hentschel}}\ and\ \bibinfo {author} {\bibfnamefont {Itamar}\ \bibnamefont
  {Procaccia}},\ }\bibfield  {title} {\enquote {\bibinfo {title} {The infinite
  number of generalized dimensions of fractals and strange attractors},}\
  }\href@noop {} {\bibfield  {journal} {\bibinfo  {journal} {Physica D:
  Nonlinear Phenomena}\ }\textbf {\bibinfo {volume} {8}},\ \bibinfo {pages}
  {435--444} (\bibinfo {year} {1983})}\BibitemShut {NoStop}%
\bibitem [{\citenamefont {Morgenstern}\ \emph {et~al.}(2003)\citenamefont
  {Morgenstern}, \citenamefont {Klijn}, \citenamefont {Meyer},\ and\
  \citenamefont {Wiesendanger}}]{morgenstern2003real}%
  \BibitemOpen
  \bibfield  {author} {\bibinfo {author} {\bibfnamefont {M}~\bibnamefont
  {Morgenstern}}, \bibinfo {author} {\bibfnamefont {J}~\bibnamefont {Klijn}},
  \bibinfo {author} {\bibfnamefont {Chr}\ \bibnamefont {Meyer}}, \ and\
  \bibinfo {author} {\bibfnamefont {R}~\bibnamefont {Wiesendanger}},\
  }\bibfield  {title} {\enquote {\bibinfo {title} {Real-space observation of
  drift states in a two-dimensional electron system at high magnetic fields},}\
  }\href@noop {} {\bibfield  {journal} {\bibinfo  {journal} {Physical Review
  Letters}\ }\textbf {\bibinfo {volume} {90}},\ \bibinfo {pages} {056804}
  (\bibinfo {year} {2003})}\BibitemShut {NoStop}%
\bibitem [{\citenamefont {Hashimoto}\ \emph {et~al.}(2008)\citenamefont
  {Hashimoto}, \citenamefont {Sohrmann}, \citenamefont {Wiebe}, \citenamefont
  {Inaoka}, \citenamefont {Meier}, \citenamefont {Hirayama}, \citenamefont
  {R{\"o}mer}, \citenamefont {Wiesendanger},\ and\ \citenamefont
  {Morgenstern}}]{hashimoto2008quantum}%
  \BibitemOpen
  \bibfield  {author} {\bibinfo {author} {\bibfnamefont {K}~\bibnamefont
  {Hashimoto}}, \bibinfo {author} {\bibfnamefont {C}~\bibnamefont {Sohrmann}},
  \bibinfo {author} {\bibfnamefont {J}~\bibnamefont {Wiebe}}, \bibinfo {author}
  {\bibfnamefont {T}~\bibnamefont {Inaoka}}, \bibinfo {author} {\bibfnamefont
  {F}~\bibnamefont {Meier}}, \bibinfo {author} {\bibfnamefont {Y}~\bibnamefont
  {Hirayama}}, \bibinfo {author} {\bibfnamefont {RA}~\bibnamefont {R{\"o}mer}},
  \bibinfo {author} {\bibfnamefont {R}~\bibnamefont {Wiesendanger}}, \ and\
  \bibinfo {author} {\bibfnamefont {M}~\bibnamefont {Morgenstern}},\ }\bibfield
   {title} {\enquote {\bibinfo {title} {Quantum hall transition in real space:
  from localized to extended states},}\ }\href@noop {} {\bibfield  {journal}
  {\bibinfo  {journal} {Physical Review Letters}\ }\textbf {\bibinfo {volume}
  {101}},\ \bibinfo {pages} {256802} (\bibinfo {year} {2008})}\BibitemShut
  {NoStop}%
\bibitem [{\citenamefont {Richardella}\ \emph {et~al.}(2010)\citenamefont
  {Richardella}, \citenamefont {Roushan}, \citenamefont {Mack}, \citenamefont
  {Zhou}, \citenamefont {Huse}, \citenamefont {Awschalom},\ and\ \citenamefont
  {Yazdani}}]{richardella2010visualizing}%
  \BibitemOpen
  \bibfield  {author} {\bibinfo {author} {\bibfnamefont {Anthony}\ \bibnamefont
  {Richardella}}, \bibinfo {author} {\bibfnamefont {Pedram}\ \bibnamefont
  {Roushan}}, \bibinfo {author} {\bibfnamefont {Shawn}\ \bibnamefont {Mack}},
  \bibinfo {author} {\bibfnamefont {Brian}\ \bibnamefont {Zhou}}, \bibinfo
  {author} {\bibfnamefont {David~A}\ \bibnamefont {Huse}}, \bibinfo {author}
  {\bibfnamefont {David~D}\ \bibnamefont {Awschalom}}, \ and\ \bibinfo {author}
  {\bibfnamefont {Ali}\ \bibnamefont {Yazdani}},\ }\bibfield  {title} {\enquote
  {\bibinfo {title} {Visualizing critical correlations near the metal-insulator
  transition in ga$_1-x$mn$_x$as},}\ }\href@noop {} {\bibfield  {journal}
  {\bibinfo  {journal} {Science}\ }\textbf {\bibinfo {volume} {327}},\ \bibinfo
  {pages} {665--669} (\bibinfo {year} {2010})}\BibitemShut {NoStop}%
\bibitem [{\citenamefont {Zhao}\ \emph {et~al.}(2019)\citenamefont {Zhao},
  \citenamefont {Lin}, \citenamefont {Xiao}, \citenamefont {Huang},
  \citenamefont {Yao}, \citenamefont {Yan}, \citenamefont {Xing}, \citenamefont
  {Zhang}, \citenamefont {Li}, \citenamefont {Hoshino} \emph
  {et~al.}}]{zhao2019disorder}%
  \BibitemOpen
  \bibfield  {author} {\bibinfo {author} {\bibfnamefont {Kun}\ \bibnamefont
  {Zhao}}, \bibinfo {author} {\bibfnamefont {Haicheng}\ \bibnamefont {Lin}},
  \bibinfo {author} {\bibfnamefont {Xiao}\ \bibnamefont {Xiao}}, \bibinfo
  {author} {\bibfnamefont {Wantong}\ \bibnamefont {Huang}}, \bibinfo {author}
  {\bibfnamefont {Wei}\ \bibnamefont {Yao}}, \bibinfo {author} {\bibfnamefont
  {Mingzhe}\ \bibnamefont {Yan}}, \bibinfo {author} {\bibfnamefont {Ying}\
  \bibnamefont {Xing}}, \bibinfo {author} {\bibfnamefont {Qinghua}\
  \bibnamefont {Zhang}}, \bibinfo {author} {\bibfnamefont {Zi-Xiang}\
  \bibnamefont {Li}}, \bibinfo {author} {\bibfnamefont {Shintaro}\ \bibnamefont
  {Hoshino}},  \emph {et~al.},\ }\bibfield  {title} {\enquote {\bibinfo {title}
  {Disorder-induced multifractal superconductivity in monolayer niobium
  dichalcogenides},}\ }\href@noop {} {\bibfield  {journal} {\bibinfo  {journal}
  {Nature Physics}\ }\textbf {\bibinfo {volume} {15}},\ \bibinfo {pages}
  {904--910} (\bibinfo {year} {2019})}\BibitemShut {NoStop}%
\bibitem [{\citenamefont {Ugeda}\ \emph {et~al.}(2016)\citenamefont {Ugeda},
  \citenamefont {Bradley}, \citenamefont {Zhang}, \citenamefont {Onishi},
  \citenamefont {Chen}, \citenamefont {Ruan}, \citenamefont
  {Ojeda-Aristizabal}, \citenamefont {Ryu}, \citenamefont {Edmonds},
  \citenamefont {Tsai} \emph {et~al.}}]{ugeda2016characterization}%
  \BibitemOpen
  \bibfield  {author} {\bibinfo {author} {\bibfnamefont {Miguel~M}\
  \bibnamefont {Ugeda}}, \bibinfo {author} {\bibfnamefont {Aaron~J}\
  \bibnamefont {Bradley}}, \bibinfo {author} {\bibfnamefont {Yi}~\bibnamefont
  {Zhang}}, \bibinfo {author} {\bibfnamefont {Seita}\ \bibnamefont {Onishi}},
  \bibinfo {author} {\bibfnamefont {Yi}~\bibnamefont {Chen}}, \bibinfo {author}
  {\bibfnamefont {Wei}\ \bibnamefont {Ruan}}, \bibinfo {author} {\bibfnamefont
  {Claudia}\ \bibnamefont {Ojeda-Aristizabal}}, \bibinfo {author}
  {\bibfnamefont {Hyejin}\ \bibnamefont {Ryu}}, \bibinfo {author}
  {\bibfnamefont {Mark~T}\ \bibnamefont {Edmonds}}, \bibinfo {author}
  {\bibfnamefont {Hsin-Zon}\ \bibnamefont {Tsai}},  \emph {et~al.},\ }\bibfield
   {title} {\enquote {\bibinfo {title} {Characterization of collective ground
  states in single-layer nbse$_2$},}\ }\href@noop {} {\bibfield  {journal}
  {\bibinfo  {journal} {Nature Physics}\ }\textbf {\bibinfo {volume} {12}},\
  \bibinfo {pages} {92--97} (\bibinfo {year} {2016})}\BibitemShut {NoStop}%
\bibitem [{\citenamefont {Bihlmayer}\ \emph {et~al.}(2007)\citenamefont
  {Bihlmayer}, \citenamefont {Bl{\"u}gel},\ and\ \citenamefont
  {Chulkov}}]{bihlmayer2007enhanced}%
  \BibitemOpen
  \bibfield  {author} {\bibinfo {author} {\bibfnamefont {Gustav}\ \bibnamefont
  {Bihlmayer}}, \bibinfo {author} {\bibfnamefont {Stefan}\ \bibnamefont
  {Bl{\"u}gel}}, \ and\ \bibinfo {author} {\bibfnamefont {EV}~\bibnamefont
  {Chulkov}},\ }\bibfield  {title} {\enquote {\bibinfo {title} {Enhanced rashba
  spin-orbit splitting in {Bi/ Ag (111) and Pb/ Ag (111)} surface alloys from
  first principles},}\ }\href@noop {} {\bibfield  {journal} {\bibinfo
  {journal} {Physical Review B}\ }\textbf {\bibinfo {volume} {75}},\ \bibinfo
  {pages} {195414} (\bibinfo {year} {2007})}\BibitemShut {NoStop}%
\bibitem [{\citenamefont {Ast}\ \emph {et~al.}(2008)\citenamefont {Ast},
  \citenamefont {Pacil{\'e}}, \citenamefont {Moreschini}, \citenamefont
  {Falub}, \citenamefont {Papagno}, \citenamefont {Kern}, \citenamefont
  {Grioni}, \citenamefont {Henk}, \citenamefont {Ernst}, \citenamefont
  {Ostanin} \emph {et~al.}}]{ast2008spin}%
  \BibitemOpen
  \bibfield  {author} {\bibinfo {author} {\bibfnamefont {Christian~R}\
  \bibnamefont {Ast}}, \bibinfo {author} {\bibfnamefont {Daniela}\ \bibnamefont
  {Pacil{\'e}}}, \bibinfo {author} {\bibfnamefont {Luca}\ \bibnamefont
  {Moreschini}}, \bibinfo {author} {\bibfnamefont {Mihaela~C}\ \bibnamefont
  {Falub}}, \bibinfo {author} {\bibfnamefont {Marco}\ \bibnamefont {Papagno}},
  \bibinfo {author} {\bibfnamefont {Klaus}\ \bibnamefont {Kern}}, \bibinfo
  {author} {\bibfnamefont {Marco}\ \bibnamefont {Grioni}}, \bibinfo {author}
  {\bibfnamefont {J{\"u}rgen}\ \bibnamefont {Henk}}, \bibinfo {author}
  {\bibfnamefont {Arthur}\ \bibnamefont {Ernst}}, \bibinfo {author}
  {\bibfnamefont {Sergey}\ \bibnamefont {Ostanin}},  \emph {et~al.},\
  }\bibfield  {title} {\enquote {\bibinfo {title} {Spin-orbit split
  two-dimensional electron gas with tunable rashba and fermi energy},}\
  }\href@noop {} {\bibfield  {journal} {\bibinfo  {journal} {Physical Review
  B}\ }\textbf {\bibinfo {volume} {77}},\ \bibinfo {pages} {081407} (\bibinfo
  {year} {2008})}\BibitemShut {NoStop}%
\bibitem [{\citenamefont {Gierz}\ \emph {et~al.}(2010)\citenamefont {Gierz},
  \citenamefont {Stadtm{\"u}ller}, \citenamefont {Vuorinen}, \citenamefont
  {Lindroos}, \citenamefont {Meier}, \citenamefont {Dil}, \citenamefont
  {Kern},\ and\ \citenamefont {Ast}}]{gierz2010structural}%
  \BibitemOpen
  \bibfield  {author} {\bibinfo {author} {\bibfnamefont {Isabella}\
  \bibnamefont {Gierz}}, \bibinfo {author} {\bibfnamefont {Benjamin}\
  \bibnamefont {Stadtm{\"u}ller}}, \bibinfo {author} {\bibfnamefont {Johannes}\
  \bibnamefont {Vuorinen}}, \bibinfo {author} {\bibfnamefont {Matti}\
  \bibnamefont {Lindroos}}, \bibinfo {author} {\bibfnamefont {Fabian}\
  \bibnamefont {Meier}}, \bibinfo {author} {\bibfnamefont {J~Hugo}\
  \bibnamefont {Dil}}, \bibinfo {author} {\bibfnamefont {Klaus}\ \bibnamefont
  {Kern}}, \ and\ \bibinfo {author} {\bibfnamefont {Christian~R}\ \bibnamefont
  {Ast}},\ }\bibfield  {title} {\enquote {\bibinfo {title} {Structural
  influence on the {Rashba-type} spin splitting in surface alloys},}\
  }\href@noop {} {\bibfield  {journal} {\bibinfo  {journal} {Physical Review
  B}\ }\textbf {\bibinfo {volume} {81}},\ \bibinfo {pages} {245430} (\bibinfo
  {year} {2010})}\BibitemShut {NoStop}%
\bibitem [{\citenamefont {Wissing}\ \emph {et~al.}(2014)\citenamefont
  {Wissing}, \citenamefont {Schmidt}, \citenamefont {Mirhosseini},
  \citenamefont {Henk}, \citenamefont {Ast},\ and\ \citenamefont
  {Donath}}]{wissing2014}%
  \BibitemOpen
  \bibfield  {author} {\bibinfo {author} {\bibfnamefont {S.~N.~P.}\
  \bibnamefont {Wissing}}, \bibinfo {author} {\bibfnamefont {A.~B.}\
  \bibnamefont {Schmidt}}, \bibinfo {author} {\bibfnamefont {H.}~\bibnamefont
  {Mirhosseini}}, \bibinfo {author} {\bibfnamefont {J.}~\bibnamefont {Henk}},
  \bibinfo {author} {\bibfnamefont {C.~R.}\ \bibnamefont {Ast}}, \ and\
  \bibinfo {author} {\bibfnamefont {M.}~\bibnamefont {Donath}},\ }\bibfield
  {title} {\enquote {\bibinfo {title} {Ambiguity of experimental spin
  information from states with mixed orbital symmetries},}\ }\href {\doibase
  10.1103/PhysRevLett.113.116402} {\bibfield  {journal} {\bibinfo  {journal}
  {Physical Review Letters}\ }\textbf {\bibinfo {volume} {113}},\ \bibinfo
  {pages} {116402} (\bibinfo {year} {2014})}\BibitemShut {NoStop}%
\bibitem [{\citenamefont {Ast}\ \emph {et~al.}(2007{\natexlab{a}})\citenamefont
  {Ast}, \citenamefont {Henk}, \citenamefont {Ernst}, \citenamefont
  {Moreschini}, \citenamefont {Falub}, \citenamefont {Pacil{\'e}},
  \citenamefont {Bruno}, \citenamefont {Kern},\ and\ \citenamefont
  {Grioni}}]{ast2007giant}%
  \BibitemOpen
  \bibfield  {author} {\bibinfo {author} {\bibfnamefont {Christian~R}\
  \bibnamefont {Ast}}, \bibinfo {author} {\bibfnamefont {J{\"u}rgen}\
  \bibnamefont {Henk}}, \bibinfo {author} {\bibfnamefont {Arthur}\ \bibnamefont
  {Ernst}}, \bibinfo {author} {\bibfnamefont {Luca}\ \bibnamefont
  {Moreschini}}, \bibinfo {author} {\bibfnamefont {Mihaela~C}\ \bibnamefont
  {Falub}}, \bibinfo {author} {\bibfnamefont {Daniela}\ \bibnamefont
  {Pacil{\'e}}}, \bibinfo {author} {\bibfnamefont {Patrick}\ \bibnamefont
  {Bruno}}, \bibinfo {author} {\bibfnamefont {Klaus}\ \bibnamefont {Kern}}, \
  and\ \bibinfo {author} {\bibfnamefont {Marco}\ \bibnamefont {Grioni}},\
  }\bibfield  {title} {\enquote {\bibinfo {title} {Giant spin splitting through
  surface alloying},}\ }\href@noop {} {\bibfield  {journal} {\bibinfo
  {journal} {Physical Review Letters}\ }\textbf {\bibinfo {volume} {98}},\
  \bibinfo {pages} {186807} (\bibinfo {year} {2007}{\natexlab{a}})}\BibitemShut
  {NoStop}%
\bibitem [{\citenamefont {Ast}\ \emph {et~al.}(2007{\natexlab{b}})\citenamefont
  {Ast}, \citenamefont {Wittich}, \citenamefont {Wahl}, \citenamefont
  {Vogelgesang}, \citenamefont {Pacil{\'e}}, \citenamefont {Falub},
  \citenamefont {Moreschini}, \citenamefont {Papagno}, \citenamefont {Grioni},\
  and\ \citenamefont {Kern}}]{ast2007local}%
  \BibitemOpen
  \bibfield  {author} {\bibinfo {author} {\bibfnamefont {Christian~R}\
  \bibnamefont {Ast}}, \bibinfo {author} {\bibfnamefont {Gero}\ \bibnamefont
  {Wittich}}, \bibinfo {author} {\bibfnamefont {Peter}\ \bibnamefont {Wahl}},
  \bibinfo {author} {\bibfnamefont {Ralf}\ \bibnamefont {Vogelgesang}},
  \bibinfo {author} {\bibfnamefont {Daniela}\ \bibnamefont {Pacil{\'e}}},
  \bibinfo {author} {\bibfnamefont {Mihaela~C}\ \bibnamefont {Falub}}, \bibinfo
  {author} {\bibfnamefont {Luca}\ \bibnamefont {Moreschini}}, \bibinfo {author}
  {\bibfnamefont {Marco}\ \bibnamefont {Papagno}}, \bibinfo {author}
  {\bibfnamefont {Marco}\ \bibnamefont {Grioni}}, \ and\ \bibinfo {author}
  {\bibfnamefont {Klaus}\ \bibnamefont {Kern}},\ }\bibfield  {title} {\enquote
  {\bibinfo {title} {Local detection of spin-orbit splitting by scanning
  tunneling spectroscopy},}\ }\href@noop {} {\bibfield  {journal} {\bibinfo
  {journal} {Physical Review B}\ }\textbf {\bibinfo {volume} {75}},\ \bibinfo
  {pages} {201401} (\bibinfo {year} {2007}{\natexlab{b}})}\BibitemShut
  {NoStop}%
\bibitem [{\citenamefont {Efros}\ and\ \citenamefont
  {Shklovskii}(1975)}]{efros1975coulomb}%
  \BibitemOpen
  \bibfield  {author} {\bibinfo {author} {\bibfnamefont {AL}~\bibnamefont
  {Efros}}\ and\ \bibinfo {author} {\bibfnamefont {Boris~I}\ \bibnamefont
  {Shklovskii}},\ }\bibfield  {title} {\enquote {\bibinfo {title} {Coulomb gap
  and low temperature conductivity of disordered systems},}\ }\href@noop {}
  {\bibfield  {journal} {\bibinfo  {journal} {Journal of Physics C: Solid State
  Physics}\ }\textbf {\bibinfo {volume} {8}},\ \bibinfo {pages} {L49} (\bibinfo
  {year} {1975})}\BibitemShut {NoStop}%
\bibitem [{\citenamefont {Altshuler}\ \emph {et~al.}(1980)\citenamefont
  {Altshuler}, \citenamefont {Aronov},\ and\ \citenamefont
  {Lee}}]{altshuler1980interaction}%
  \BibitemOpen
  \bibfield  {author} {\bibinfo {author} {\bibfnamefont {Boris~L}\ \bibnamefont
  {Altshuler}}, \bibinfo {author} {\bibfnamefont {Arkadi~G}\ \bibnamefont
  {Aronov}}, \ and\ \bibinfo {author} {\bibfnamefont {PA}~\bibnamefont {Lee}},\
  }\bibfield  {title} {\enquote {\bibinfo {title} {Interaction effects in
  disordered fermi systems in two dimensions},}\ }\href@noop {} {\bibfield
  {journal} {\bibinfo  {journal} {Physical Review Letters}\ }\textbf {\bibinfo
  {volume} {44}},\ \bibinfo {pages} {1288} (\bibinfo {year}
  {1980})}\BibitemShut {NoStop}%
\bibitem [{\citenamefont {Imry}\ and\ \citenamefont
  {Ovadyahu}(1982)}]{imry1982density}%
  \BibitemOpen
  \bibfield  {author} {\bibinfo {author} {\bibfnamefont {Yoseph}\ \bibnamefont
  {Imry}}\ and\ \bibinfo {author} {\bibfnamefont {Zvi}\ \bibnamefont
  {Ovadyahu}},\ }\bibfield  {title} {\enquote {\bibinfo {title}
  {Density-of-states anomalies in a disordered conductor: a tunneling study},}\
  }\href@noop {} {\bibfield  {journal} {\bibinfo  {journal} {Physical Review
  Letters}\ }\textbf {\bibinfo {volume} {49}},\ \bibinfo {pages} {841}
  (\bibinfo {year} {1982})}\BibitemShut {NoStop}%
\bibitem [{\citenamefont {Bielejec}\ \emph {et~al.}(2001)\citenamefont
  {Bielejec}, \citenamefont {Ruan},\ and\ \citenamefont
  {Wu}}]{bielejec2001hard}%
  \BibitemOpen
  \bibfield  {author} {\bibinfo {author} {\bibfnamefont {E}~\bibnamefont
  {Bielejec}}, \bibinfo {author} {\bibfnamefont {J}~\bibnamefont {Ruan}}, \
  and\ \bibinfo {author} {\bibfnamefont {Wenhao}\ \bibnamefont {Wu}},\
  }\bibfield  {title} {\enquote {\bibinfo {title} {Hard correlation gap
  observed in quench-condensed ultrathin beryllium},}\ }\href@noop {}
  {\bibfield  {journal} {\bibinfo  {journal} {Physical Review Letters}\
  }\textbf {\bibinfo {volume} {87}},\ \bibinfo {pages} {036801} (\bibinfo
  {year} {2001})}\BibitemShut {NoStop}%
\bibitem [{\citenamefont {Ando}(1989)}]{ando1989numerical}%
  \BibitemOpen
  \bibfield  {author} {\bibinfo {author} {\bibfnamefont {T}~\bibnamefont
  {Ando}},\ }\bibfield  {title} {\enquote {\bibinfo {title} {Numerical study of
  symmetry effects on localization in two dimensions},}\ }\href@noop {}
  {\bibfield  {journal} {\bibinfo  {journal} {Physical Review B}\ }\textbf
  {\bibinfo {volume} {40}},\ \bibinfo {pages} {5325} (\bibinfo {year}
  {1989})}\BibitemShut {NoStop}%
\bibitem [{\citenamefont {Mildenberger}\ and\ \citenamefont
  {Evers}(2007)}]{mildenberger2007wave}%
  \BibitemOpen
  \bibfield  {author} {\bibinfo {author} {\bibfnamefont {A}~\bibnamefont
  {Mildenberger}}\ and\ \bibinfo {author} {\bibfnamefont {F}~\bibnamefont
  {Evers}},\ }\bibfield  {title} {\enquote {\bibinfo {title} {Wave function
  statistics at the symplectic two-dimensional anderson transition: Bulk
  properties},}\ }\href@noop {} {\bibfield  {journal} {\bibinfo  {journal}
  {Physical Review B}\ }\textbf {\bibinfo {volume} {75}},\ \bibinfo {pages}
  {041303} (\bibinfo {year} {2007})}\BibitemShut {NoStop}%
\bibitem [{\citenamefont {Hikami}\ \emph {et~al.}(1980)\citenamefont {Hikami},
  \citenamefont {Larkin},\ and\ \citenamefont {Nagaoka}}]{hikami1980spin}%
  \BibitemOpen
  \bibfield  {author} {\bibinfo {author} {\bibfnamefont {Shinobu}\ \bibnamefont
  {Hikami}}, \bibinfo {author} {\bibfnamefont {Anatoly~I}\ \bibnamefont
  {Larkin}}, \ and\ \bibinfo {author} {\bibfnamefont {Yosuke}\ \bibnamefont
  {Nagaoka}},\ }\bibfield  {title} {\enquote {\bibinfo {title} {Spin-orbit
  interaction and magnetoresistance in the two dimensional random system},}\
  }\href@noop {} {\bibfield  {journal} {\bibinfo  {journal} {Progress of
  Theoretical Physics}\ }\textbf {\bibinfo {volume} {63}},\ \bibinfo {pages}
  {707--710} (\bibinfo {year} {1980})}\BibitemShut {NoStop}%
\bibitem [{\citenamefont {Lifshitz}\ \emph {et~al.}(1965)\citenamefont
  {Lifshitz} \emph {et~al.}}]{lifshitz1965energy}%
  \BibitemOpen
  \bibfield  {author} {\bibinfo {author} {\bibfnamefont {Il'ya~Mikhailovich}\
  \bibnamefont {Lifshitz}} \emph {et~al.},\ }\bibfield  {title} {\enquote
  {\bibinfo {title} {Energy spectrum structure and quantum states of disordered
  condensed systems},}\ }\href@noop {} {\bibfield  {journal} {\bibinfo
  {journal} {Soviet Physics Uspekhi}\ }\textbf {\bibinfo {volume} {7}},\
  \bibinfo {pages} {549} (\bibinfo {year} {1965})}\BibitemShut {NoStop}%
\bibitem [{\citenamefont {Mott}(1968)}]{mott1968metal}%
  \BibitemOpen
  \bibfield  {author} {\bibinfo {author} {\bibfnamefont {NF}~\bibnamefont
  {Mott}},\ }\bibfield  {title} {\enquote {\bibinfo {title} {Metal-insulator
  transition},}\ }\href@noop {} {\bibfield  {journal} {\bibinfo  {journal}
  {Reviews of Modern Physics}\ }\textbf {\bibinfo {volume} {40}},\ \bibinfo
  {pages} {677} (\bibinfo {year} {1968})}\BibitemShut {NoStop}%
\bibitem [{\citenamefont {Wegner}(1980)}]{wegner1980inverse}%
  \BibitemOpen
  \bibfield  {author} {\bibinfo {author} {\bibfnamefont {Franz}\ \bibnamefont
  {Wegner}},\ }\bibfield  {title} {\enquote {\bibinfo {title} {Inverse
  participation ratio in 2+ $\varepsilon$ dimensions},}\ }\href@noop {}
  {\bibfield  {journal} {\bibinfo  {journal} {Zeitschrift f{\"u}r Physik B
  Condensed Matter}\ }\textbf {\bibinfo {volume} {36}},\ \bibinfo {pages}
  {209--214} (\bibinfo {year} {1980})}\BibitemShut {NoStop}%
\bibitem [{\citenamefont {Chhabra}\ and\ \citenamefont
  {Jensen}(1989)}]{chhabra1989direct}%
  \BibitemOpen
  \bibfield  {author} {\bibinfo {author} {\bibfnamefont {Ashvin}\ \bibnamefont
  {Chhabra}}\ and\ \bibinfo {author} {\bibfnamefont {Roderick~V}\ \bibnamefont
  {Jensen}},\ }\bibfield  {title} {\enquote {\bibinfo {title} {Direct
  determination of the f ($\alpha$) singularity spectrum},}\ }\href@noop {}
  {\bibfield  {journal} {\bibinfo  {journal} {Physical Review Letters}\
  }\textbf {\bibinfo {volume} {62}},\ \bibinfo {pages} {1327} (\bibinfo {year}
  {1989})}\BibitemShut {NoStop}%
\bibitem [{\citenamefont {Mirlin}\ \emph {et~al.}(2006)\citenamefont {Mirlin},
  \citenamefont {Fyodorov}, \citenamefont {Mildenberger},\ and\ \citenamefont
  {Evers}}]{mirlin2006exact}%
  \BibitemOpen
  \bibfield  {author} {\bibinfo {author} {\bibfnamefont {AD}~\bibnamefont
  {Mirlin}}, \bibinfo {author} {\bibfnamefont {Yu~V}\ \bibnamefont {Fyodorov}},
  \bibinfo {author} {\bibfnamefont {A}~\bibnamefont {Mildenberger}}, \ and\
  \bibinfo {author} {\bibfnamefont {F}~\bibnamefont {Evers}},\ }\bibfield
  {title} {\enquote {\bibinfo {title} {Exact relations between multifractal
  exponents at the anderson transition},}\ }\href@noop {} {\bibfield  {journal}
  {\bibinfo  {journal} {Physical Review Letters}\ }\textbf {\bibinfo {volume}
  {97}},\ \bibinfo {pages} {046803} (\bibinfo {year} {2006})}\BibitemShut
  {NoStop}%
\bibitem [{\citenamefont {Rubio-Verd{\'u}}\ \emph {et~al.}(2020)\citenamefont
  {Rubio-Verd{\'u}}, \citenamefont {Garc{\'\i}a-Garc{\'\i}a}, \citenamefont
  {Ryu}, \citenamefont {Choi}, \citenamefont {Zald{\'\i}var}, \citenamefont
  {Tang}, \citenamefont {Fan}, \citenamefont {Shen}, \citenamefont {Mo},
  \citenamefont {Pascual},\ and\ \citenamefont
  {Ugeda}}]{rubio2018visualization}%
  \BibitemOpen
  \bibfield  {author} {\bibinfo {author} {\bibfnamefont {Carmen}\ \bibnamefont
  {Rubio-Verd{\'u}}}, \bibinfo {author} {\bibfnamefont {Antonio~M.}\
  \bibnamefont {Garc{\'\i}a-Garc{\'\i}a}}, \bibinfo {author} {\bibfnamefont
  {Hyejin}\ \bibnamefont {Ryu}}, \bibinfo {author} {\bibfnamefont {Deung-Jang}\
  \bibnamefont {Choi}}, \bibinfo {author} {\bibfnamefont {Javier}\ \bibnamefont
  {Zald{\'\i}var}}, \bibinfo {author} {\bibfnamefont {Shujie}\ \bibnamefont
  {Tang}}, \bibinfo {author} {\bibfnamefont {Bo}~\bibnamefont {Fan}}, \bibinfo
  {author} {\bibfnamefont {Zhi-Xun}\ \bibnamefont {Shen}}, \bibinfo {author}
  {\bibfnamefont {Sung-Kwan}\ \bibnamefont {Mo}}, \bibinfo {author}
  {\bibfnamefont {Jos{\'e}Ignacio}\ \bibnamefont {Pascual}}, \ and\ \bibinfo
  {author} {\bibfnamefont {Miguel~M.}\ \bibnamefont {Ugeda}},\ }\bibfield
  {title} {\enquote {\bibinfo {title} {Visualization of multifractal
  superconductivity in a two-dimensional transition metal dichalcogenide in the
  weak-disorder regime},}\ }\bibfield  {booktitle} {\emph {\bibinfo {booktitle}
  {Nano Letters}},\ }\href {\doibase 10.1021/acs.nanolett.0c01288} {\bibfield
  {journal} {\bibinfo  {journal} {Nano Letters}\ }\textbf {\bibinfo {volume}
  {20}},\ \bibinfo {pages} {5111--5118} (\bibinfo {year} {2020})}\BibitemShut
  {NoStop}%
\bibitem [{\citenamefont {Feigel\'{}man}\ \emph {et~al.}(2007)\citenamefont
  {Feigel\'{}man}, \citenamefont {Ioffe}, \citenamefont {Kravtsov},\ and\
  \citenamefont {Yuzbashyan}}]{feigel2007eigenfunction}%
  \BibitemOpen
  \bibfield  {author} {\bibinfo {author} {\bibfnamefont {MV}~\bibnamefont
  {Feigel\'{}man}}, \bibinfo {author} {\bibfnamefont {LB}~\bibnamefont
  {Ioffe}}, \bibinfo {author} {\bibfnamefont {VE}~\bibnamefont {Kravtsov}}, \
  and\ \bibinfo {author} {\bibfnamefont {EA}~\bibnamefont {Yuzbashyan}},\
  }\bibfield  {title} {\enquote {\bibinfo {title} {Eigenfunction fractality and
  pseudogap state near the superconductor-insulator transition},}\ }\href@noop
  {} {\bibfield  {journal} {\bibinfo  {journal} {Physical Review Letters}\
  }\textbf {\bibinfo {volume} {98}},\ \bibinfo {pages} {027001} (\bibinfo
  {year} {2007})}\BibitemShut {NoStop}%
\bibitem [{\citenamefont {Feigel'man}\ \emph {et~al.}(2010)\citenamefont
  {Feigel'man}, \citenamefont {Ioffe}, \citenamefont {Kravtsov},\ and\
  \citenamefont {Cuevas}}]{feigel2010fractal}%
  \BibitemOpen
  \bibfield  {author} {\bibinfo {author} {\bibfnamefont {MV}~\bibnamefont
  {Feigel'man}}, \bibinfo {author} {\bibfnamefont {LB}~\bibnamefont {Ioffe}},
  \bibinfo {author} {\bibfnamefont {VE}~\bibnamefont {Kravtsov}}, \ and\
  \bibinfo {author} {\bibfnamefont {E}~\bibnamefont {Cuevas}},\ }\bibfield
  {title} {\enquote {\bibinfo {title} {Fractal superconductivity near
  localization threshold},}\ }\href@noop {} {\bibfield  {journal} {\bibinfo
  {journal} {Annals of Physics}\ }\textbf {\bibinfo {volume} {325}},\ \bibinfo
  {pages} {1390--1478} (\bibinfo {year} {2010})}\BibitemShut {NoStop}%
\bibitem [{\citenamefont {Burmistrov}\ \emph {et~al.}(2012)\citenamefont
  {Burmistrov}, \citenamefont {Gornyi},\ and\ \citenamefont
  {Mirlin}}]{burmistrov2012enhancement}%
  \BibitemOpen
  \bibfield  {author} {\bibinfo {author} {\bibfnamefont {IS}~\bibnamefont
  {Burmistrov}}, \bibinfo {author} {\bibfnamefont {IV}~\bibnamefont {Gornyi}},
  \ and\ \bibinfo {author} {\bibfnamefont {AD}~\bibnamefont {Mirlin}},\
  }\bibfield  {title} {\enquote {\bibinfo {title} {Enhancement of the critical
  temperature of superconductors by anderson localization},}\ }\href@noop {}
  {\bibfield  {journal} {\bibinfo  {journal} {Physical Review Letters}\
  }\textbf {\bibinfo {volume} {108}},\ \bibinfo {pages} {017002} (\bibinfo
  {year} {2012})}\BibitemShut {NoStop}%
\bibitem [{\citenamefont {Mayoh}\ and\ \citenamefont
  {Garc{\'\i}a-Garc{\'\i}a}(2015)}]{mayoh2015global}%
  \BibitemOpen
  \bibfield  {author} {\bibinfo {author} {\bibfnamefont {James}\ \bibnamefont
  {Mayoh}}\ and\ \bibinfo {author} {\bibfnamefont {Antonio~M}\ \bibnamefont
  {Garc{\'\i}a-Garc{\'\i}a}},\ }\bibfield  {title} {\enquote {\bibinfo {title}
  {Global critical temperature in disordered superconductors with weak
  multifractality},}\ }\href@noop {} {\bibfield  {journal} {\bibinfo  {journal}
  {Physical Review B}\ }\textbf {\bibinfo {volume} {92}},\ \bibinfo {pages}
  {174526} (\bibinfo {year} {2015})}\BibitemShut {NoStop}%
\bibitem [{\citenamefont {Ko{\v{z}}elj}\ \emph {et~al.}(2014)\citenamefont
  {Ko{\v{z}}elj}, \citenamefont {Vrtnik}, \citenamefont {Jelen}, \citenamefont
  {Jazbec}, \citenamefont {Jagli{\v{c}}i{\'c}}, \citenamefont {Maiti},
  \citenamefont {Feuerbacher}, \citenamefont {Steurer},\ and\ \citenamefont
  {Dolin{\v{s}}ek}}]{kovzelj2014discovery}%
  \BibitemOpen
  \bibfield  {author} {\bibinfo {author} {\bibfnamefont {Primo{\v{z}}}\
  \bibnamefont {Ko{\v{z}}elj}}, \bibinfo {author} {\bibfnamefont {Stanislav}\
  \bibnamefont {Vrtnik}}, \bibinfo {author} {\bibfnamefont {Andreja}\
  \bibnamefont {Jelen}}, \bibinfo {author} {\bibfnamefont {Simon}\ \bibnamefont
  {Jazbec}}, \bibinfo {author} {\bibfnamefont {Zvonko}\ \bibnamefont
  {Jagli{\v{c}}i{\'c}}}, \bibinfo {author} {\bibfnamefont {S}~\bibnamefont
  {Maiti}}, \bibinfo {author} {\bibfnamefont {Michael}\ \bibnamefont
  {Feuerbacher}}, \bibinfo {author} {\bibfnamefont {Walter}\ \bibnamefont
  {Steurer}}, \ and\ \bibinfo {author} {\bibfnamefont {Janez}\ \bibnamefont
  {Dolin{\v{s}}ek}},\ }\bibfield  {title} {\enquote {\bibinfo {title}
  {Discovery of a superconducting high-entropy alloy},}\ }\href@noop {}
  {\bibfield  {journal} {\bibinfo  {journal} {Physical Review Letters}\
  }\textbf {\bibinfo {volume} {113}},\ \bibinfo {pages} {107001} (\bibinfo
  {year} {2014})}\BibitemShut {NoStop}%
\bibitem [{\citenamefont {Sun}\ and\ \citenamefont {Cava}(2019)}]{sun2019high}%
  \BibitemOpen
  \bibfield  {author} {\bibinfo {author} {\bibfnamefont {Liling}\ \bibnamefont
  {Sun}}\ and\ \bibinfo {author} {\bibfnamefont {RJ}~\bibnamefont {Cava}},\
  }\bibfield  {title} {\enquote {\bibinfo {title} {High-entropy alloy
  superconductors: Status, opportunities, and challenges},}\ }\href@noop {}
  {\bibfield  {journal} {\bibinfo  {journal} {Physical Review Materials}\
  }\textbf {\bibinfo {volume} {3}},\ \bibinfo {pages} {090301} (\bibinfo {year}
  {2019})}\BibitemShut {NoStop}%
\bibitem [{\citenamefont {Budke}\ \emph {et~al.}(2007)\citenamefont {Budke},
  \citenamefont {Renken}, \citenamefont {Liebl}, \citenamefont {Rangelov},\
  and\ \citenamefont {Donath}}]{budke2007inverse}%
  \BibitemOpen
  \bibfield  {author} {\bibinfo {author} {\bibfnamefont {M}~\bibnamefont
  {Budke}}, \bibinfo {author} {\bibfnamefont {V}~\bibnamefont {Renken}},
  \bibinfo {author} {\bibfnamefont {H}~\bibnamefont {Liebl}}, \bibinfo {author}
  {\bibfnamefont {G}~\bibnamefont {Rangelov}}, \ and\ \bibinfo {author}
  {\bibfnamefont {M}~\bibnamefont {Donath}},\ }\bibfield  {title} {\enquote
  {\bibinfo {title} {Inverse photoemission with energy resolution better than
  200 mev},}\ }\href@noop {} {\bibfield  {journal} {\bibinfo  {journal} {Review
  of Scientific Instruments}\ }\textbf {\bibinfo {volume} {78}},\ \bibinfo
  {pages} {083903} (\bibinfo {year} {2007})}\BibitemShut {NoStop}%
\bibitem [{\citenamefont {Thiede}\ \emph {et~al.}(2018)\citenamefont {Thiede},
  \citenamefont {Niehues}, \citenamefont {Schmidt},\ and\ \citenamefont
  {Donath}}]{thiede2018acetone}%
  \BibitemOpen
  \bibfield  {author} {\bibinfo {author} {\bibfnamefont {Christian}\
  \bibnamefont {Thiede}}, \bibinfo {author} {\bibfnamefont {Iris}\ \bibnamefont
  {Niehues}}, \bibinfo {author} {\bibfnamefont {Anke~B}\ \bibnamefont
  {Schmidt}}, \ and\ \bibinfo {author} {\bibfnamefont {Markus}\ \bibnamefont
  {Donath}},\ }\bibfield  {title} {\enquote {\bibinfo {title} {The acetone
  bandpass detector for inverse photoemission: operation in proportional and
  geiger--m{\"u}ller modes},}\ }\href@noop {} {\bibfield  {journal} {\bibinfo
  {journal} {Measurement Science and Technology}\ }\textbf {\bibinfo {volume}
  {29}},\ \bibinfo {pages} {065901} (\bibinfo {year} {2018})}\BibitemShut
  {NoStop}%
\bibitem [{\citenamefont {Zumb{\"u}lte}\ \emph {et~al.}(2015)\citenamefont
  {Zumb{\"u}lte}, \citenamefont {Schmidt},\ and\ \citenamefont
  {Donath}}]{zumbulte2015momentum}%
  \BibitemOpen
  \bibfield  {author} {\bibinfo {author} {\bibfnamefont {A}~\bibnamefont
  {Zumb{\"u}lte}}, \bibinfo {author} {\bibfnamefont {AB}~\bibnamefont
  {Schmidt}}, \ and\ \bibinfo {author} {\bibfnamefont {M}~\bibnamefont
  {Donath}},\ }\bibfield  {title} {\enquote {\bibinfo {title} {Momentum
  resolution in inverse photoemission},}\ }\href@noop {} {\bibfield  {journal}
  {\bibinfo  {journal} {Review of Scientific Instruments}\ }\textbf {\bibinfo
  {volume} {86}},\ \bibinfo {pages} {013908} (\bibinfo {year}
  {2015})}\BibitemShut {NoStop}%
\bibitem [{\citenamefont {Alvermann}\ and\ \citenamefont
  {Fehske}(2005)}]{alvermann2005local}%
  \BibitemOpen
  \bibfield  {author} {\bibinfo {author} {\bibfnamefont {Andreas}\ \bibnamefont
  {Alvermann}}\ and\ \bibinfo {author} {\bibfnamefont {Holger}\ \bibnamefont
  {Fehske}},\ }\bibfield  {title} {\enquote {\bibinfo {title} {Local
  distribution approach to disordered binary alloys},}\ }\href@noop {}
  {\bibfield  {journal} {\bibinfo  {journal} {The European Physical Journal
  B-Condensed Matter and Complex Systems}\ }\textbf {\bibinfo {volume} {48}},\
  \bibinfo {pages} {295--303} (\bibinfo {year} {2005})}\BibitemShut {NoStop}%
\bibitem [{\citenamefont {Schubert}\ \emph {et~al.}(2010)\citenamefont
  {Schubert}, \citenamefont {Schleede}, \citenamefont {Byczuk}, \citenamefont
  {Fehske},\ and\ \citenamefont {Vollhardt}}]{schubert2010distribution}%
  \BibitemOpen
  \bibfield  {author} {\bibinfo {author} {\bibfnamefont {Gerald}\ \bibnamefont
  {Schubert}}, \bibinfo {author} {\bibfnamefont {Jens}\ \bibnamefont
  {Schleede}}, \bibinfo {author} {\bibfnamefont {Krzysztof}\ \bibnamefont
  {Byczuk}}, \bibinfo {author} {\bibfnamefont {Holger}\ \bibnamefont {Fehske}},
  \ and\ \bibinfo {author} {\bibfnamefont {Dieter}\ \bibnamefont {Vollhardt}},\
  }\bibfield  {title} {\enquote {\bibinfo {title} {Distribution of the local
  density of states as a criterion for anderson localization: Numerically exact
  results for various lattices in two and three dimensions},}\ }\href@noop {}
  {\bibfield  {journal} {\bibinfo  {journal} {Physical Review B}\ }\textbf
  {\bibinfo {volume} {81}},\ \bibinfo {pages} {155106} (\bibinfo {year}
  {2010})}\BibitemShut {NoStop}%
\bibitem [{\citenamefont {Asada}\ \emph {et~al.}(2002)\citenamefont {Asada},
  \citenamefont {Slevin},\ and\ \citenamefont {Ohtsuki}}]{asada2002anderson}%
  \BibitemOpen
  \bibfield  {author} {\bibinfo {author} {\bibfnamefont {Yoichi}\ \bibnamefont
  {Asada}}, \bibinfo {author} {\bibfnamefont {Keith}\ \bibnamefont {Slevin}}, \
  and\ \bibinfo {author} {\bibfnamefont {Tomi}\ \bibnamefont {Ohtsuki}},\
  }\bibfield  {title} {\enquote {\bibinfo {title} {Anderson transition in
  two-dimensional systems with spin-orbit coupling},}\ }\href@noop {}
  {\bibfield  {journal} {\bibinfo  {journal} {Physical Review Letters}\
  }\textbf {\bibinfo {volume} {89}},\ \bibinfo {pages} {256601} (\bibinfo
  {year} {2002})}\BibitemShut {NoStop}%
\bibitem [{\citenamefont {Ku}\ \emph {et~al.}(2010)\citenamefont {Ku},
  \citenamefont {Berlijn}, \citenamefont {Lee} \emph
  {et~al.}}]{ku2010unfolding}%
  \BibitemOpen
  \bibfield  {author} {\bibinfo {author} {\bibfnamefont {Wei}\ \bibnamefont
  {Ku}}, \bibinfo {author} {\bibfnamefont {Tom}\ \bibnamefont {Berlijn}},
  \bibinfo {author} {\bibfnamefont {Chi-Cheng}\ \bibnamefont {Lee}},  \emph
  {et~al.},\ }\bibfield  {title} {\enquote {\bibinfo {title} {Unfolding
  first-principles band structures},}\ }\href@noop {} {\bibfield  {journal}
  {\bibinfo  {journal} {Physical Review Letters}\ }\textbf {\bibinfo {volume}
  {104}},\ \bibinfo {pages} {216401} (\bibinfo {year} {2010})}\BibitemShut
  {NoStop}%
\bibitem [{\citenamefont {Popescu}\ and\ \citenamefont
  {Zunger}(2010)}]{popescu2010effective}%
  \BibitemOpen
  \bibfield  {author} {\bibinfo {author} {\bibfnamefont {Voicu}\ \bibnamefont
  {Popescu}}\ and\ \bibinfo {author} {\bibfnamefont {Alex}\ \bibnamefont
  {Zunger}},\ }\bibfield  {title} {\enquote {\bibinfo {title} {Effective band
  structure of random alloys},}\ }\href@noop {} {\bibfield  {journal} {\bibinfo
   {journal} {Physical Review Letters}\ }\textbf {\bibinfo {volume} {104}},\
  \bibinfo {pages} {236403} (\bibinfo {year} {2010})}\BibitemShut {NoStop}%
\bibitem [{\citenamefont {Haverkort}\ \emph {et~al.}(2011)\citenamefont
  {Haverkort}, \citenamefont {Elfimov},\ and\ \citenamefont
  {Sawatzky}}]{haverkort2011electronic}%
  \BibitemOpen
  \bibfield  {author} {\bibinfo {author} {\bibfnamefont {MW}~\bibnamefont
  {Haverkort}}, \bibinfo {author} {\bibfnamefont {IS}~\bibnamefont {Elfimov}},
  \ and\ \bibinfo {author} {\bibfnamefont {GA}~\bibnamefont {Sawatzky}},\
  }\bibfield  {title} {\enquote {\bibinfo {title} {Electronic structure and
  self energies of randomly substituted solids using density functional theory
  and model calculations},}\ }\href@noop {} {\bibfield  {journal} {\bibinfo
  {journal} {arXiv preprint arXiv:1109.4036}\ } (\bibinfo {year}
  {2011})}\BibitemShut {NoStop}%
\end{thebibliography}%

\end{document}